\documentclass[12pt]{article}
\usepackage{booktabs}
\usepackage{epsfig, amssymb}
\usepackage{amsfonts}
\usepackage{simpler-wick}
\usepackage{amsmath}
\usepackage{booktabs, colortbl, array}
\usepackage{float}
\usepackage{graphicx,epsfig,caption,dblfloatfix}
\usepackage{color}
\usepackage{cite}
\begin{document}
\thispagestyle{empty}

\begin{flushright}
\end{flushright}

\bigskip
\begin{center}
\noindent{\Large \textbf
{Analytic approaches to anisotropic holographic superfluids in asymptotically hyperscaling violation geometry}}\\ 

\vspace{2cm}  
\noindent{Gitae Kim${}^{a}$\footnote{email : kimgitae728@gmail.com}, Yeong Seok Choi${}^{a}$\footnote{email : sehhearcoatz@gmail.com} and Jae-Hyuk Oh${}^{a}$\footnote{email : jaehyukoh@hanyang.ac.kr}}

\vspace{1cm}
  {\it
Department of Physics, Hanyang University, Seoul 04763, Korea${}^{a}$\\
 }
\end{center}

\vspace{0.3cm}
\begin{abstract}
\noindent
We explore anisotropic holographic superfluidity and find analytic solutions near critical point where superfluid/normalfluid phase transition appears in a holographic dual fluid system. In arXiv:hep-th/1109.4592, the authors obtained such an analytic solution in 5-dimensional Einstein-SU(2)Yang-Mills system, and it is asymptotically AdS$_5$. What is more in this note is that we get analytic solutions near the critical point in 3- and 4- dimensional Einstein-Scalar-U(1)$\times$SU(2)Yang-Mills systems, which become asymptotically hyperscaling violation geometry. We also get leading order back reactions to the background geometry, which clearly shows spatial anisotropy of the bulk geometry. To explore the properties of the spacetime, we compute holographic entanglement entropy and confirm that the superfluid/normalfluid phase transition must occur at the critical point.

\end{abstract}

\newpage
\vspace{0.3cm}
\noindent
\tableofcontents

\section{Introduction}
	\label{section_introduction}
Strongly interacting systems are very hard to explore since the usual perturbation theory does not work in there.
One of the noticeable methodology to study such strongly interacting physics is gauge/gravity duality\cite{Aharony:1999ti,Maldacena:1997re,Oh:2011wpl}.
This duality applies to (conformal) fluid dynamics\cite{ Kovtun:2004de,Bhattacharyya:2007vjd,Bhattacharyya:2008ji,Benincasa:2006fu,Iqbal:2008by}, condensed matter systems\cite{Hartnoll:2009sz,Gubser:2008px,Gubser:2008zu,Hartnoll:2008kx,Basu:2008bh,Herzog:2009xv,Horowitz:2010gk} and so on, by employing their corresponding dual gravity systems and contribute to explore those systems in strongly interacting regime.
Among them, application of holography to conformal fluid dynamics was quite successful and one of the example that we should stress is the holographic computation of the ratio of shear viscosity to entropy density\cite{Kovtun:2004de,Bhattacharyya:2007vjd,Bhattacharyya:2008ji,Benincasa:2006fu,Iqbal:2008by,Erdmenger:2010xm,Basu:2011tt,Oh:2012zu}. 

The studies on holographic superfluid(superconductor) are started with a simple gravity model, Einstein-U(1)-complex scalar model\cite{Hartnoll:2008kx}.
In this model, above a certain critical temperature $T_c$, the gravity system is just a charged black brane solution.
However, below $T_c$, a non-trivial scalar hair appears and it is an indication of symmetry breaking in the dual conformal fluid(condensed matter) systems.
Such a gravity system is called s-wave holographic superfluid(condensed matter) systems.

The s-wave holographic fluid model has been extended to p-wave superfluid(superconductor) models, where the symmetry breaking order parameter appearing in the dual field theory system is vector-like. These models have been extensively studied in\cite{Herzog:2009ci,Basu:2008bh,Arias:2012py,Cai:2012nm}.
Especially, in the previous studies\cite{Basu:2011tt,Oh:2012zu,Park:2022oek}, the authors devoted to holographic dual gravity model of Einstein-SU(2)Yang-Mills theory in asymptotically AdS$_5$ and they found thermodynamic phase transition from isotropic to anisotropic phases at the (rescaled and so dimensionless)chemical potential $\mu=4/\sqrt{3}$.
What is interesting in these studies is that at this value of chemical potential, there is an analytic solution of the spatial component of SU(2)Yang-Mills field which breaks $O(3)$-global rotation symmetry down to $O(2)$.
We note that this Yang-Mills solution is obtained by solving Yang-Mills field equations in the given AdS background.
In this case, $\epsilon=\frac{1}{2} g^2_{YM} \langle j^1_x \rangle$ appears as the order parameter, where $\langle j^1_x \rangle$ is the expectation value of the non-abelian current.

Moreover, the leading backreaction of the anisotropic solution of Yang-Mills field to the background spacetime is obtained in \cite{Basu:2011tt}. They compute leading order backreactions of the Yang-Mills field to the background spacetime, and do compute directional dependence of the ratio of shear viscosity to entropy density in dual fluid system. The holographic renormalization group flows of the ratio is discussed by using the analytic solutions in \cite{Oh:2012zu}. There are also analytic calculations of holographic entanglement entropy computations in the anisotropic background in \cite{Park:2022oek}.

In \cite{Park:2016wch}, the authors extended the p-wave superfluid models to asymptotically Lifshitz and hyperscaling violation geometries. They consider Einstein-Scalar-U(1)$\times$SU(2)Yang-Mills systems in 5-dimensional spacetime, where the scalar and U(1) gauge fields have roles which deform the asymptotic boundary to become Lifshitz and hyperscaling violation. The SU(2) gauge field has the same role with the previous cases, which provides chemical potential and causes  phase transition to anisotripic phase from isotropic phase, where the vector order appears at a certain critical value of the chemical potential.
Some of conditions are imposed on $d, z$ and $\alpha$, which are the theory parameters given in the gravity model together with the null energy condition, where $d$ is the spatial dimension of boundary spacetime ($d=D-2$), $z$ the Lifshitz scaling factor, and $\alpha$ the hyperscaling violation factor. $D$ is the dimensionality of the bulk space and time where the gravity theory is defined in.


In this paper, we find analytic solutions for 3- and 4-dimensional asymptotically hyperscaling violating black brane, charged by U(1)$\times$SU(2) gauge fields.
We obtain analytic solutions for the Yang-Mills fields in the Einstein-Scalar-U(1)$\times$SU(2) theory, by choosing an ansatz which represents a vector order in $x_1$ direction.
We choose $z=1$, then the null energy condition restricts the value of $\alpha=2$ for $D=3$ case and $\alpha=\frac{1}{2}$ for $D=4$ case.
It turns out that such solutions are possible when the rescaled dimensionless chemical potential $\mu=4/\sqrt{3}$.

Moreover, we consider backreactions from the Yang-Mills fields to the background fields including bulk spacetime.
We solve the equations of motion of the gravity system with power expansion order by order in small parameters of $\epsilon$ as well as $\delta=\frac{\kappa_D}{g_{YM}}r_0 \ll 1$, where $\kappa_D$ is $D$-dimensional gravity constant and $g_{YM}$ is Yang-Mills coupling constant and $r_0$ is the blackbrane horizon.
We note that both $\epsilon$ and $\delta$ are dimensionless.
We compute the backreactions up to $\epsilon^2\delta^2$-order terms, which is the leading order.
We leave the background spacetime to have the horizon at $u=1$ and be asymptotically hyperscaling violating at boundary, where $u$ is the rescaled radial coordinate given by $u=r/r_0$ and $r_0$ is the horizon in the original coordinate.

Because the vector order breaks the global isometry, we expect that there exists a phase transition from isotropic to anisotropic phase.
One of the ways to probe this system is computing (holographic) entanglement entropy.
With the analytic solutions at hand, we calculate holographic entanglement entropies of some subsystems defined on asymptotic boundary.
The entanglement entropy of a system is obtained by minimizing its surface area stretched down into the bulk\cite{Ryu:2006bv,Ryu:2006ef}.

For $D=3$ case, we consider a system whose subsystem at boundary space is a line segment.
For $D=4$ case, we consider two systems whose subsystems at boundary space have the shape of a strap, but one is along the direction of the vector order, and another is normal to the vector order.
We compute the holographic entanglement entropy in the regime of small subsystem on the asymptotic boundary so that one can construct the entanglement by employing perturbation with small subsystem size.

It turns out that the entanglement entropies of the three systems are corrected by leading corrections being order of $\epsilon^2\delta^2$-terms.
It indicates that a phase transition occurs when the vector order appears.
In fact, the holographic entanglement entropy of isotropic phase is smaller than that of anisotropic phase.
Therefore the systems prefer the anisotropic phase over the isotropic one.

We can slso see that the first law of entanglement entropy is valid for the three systems, by showing that the energy and entanglement entropy changes are proportional to each other.
The entanglement temperature $\mathcal{T}=\Delta E / \Delta S$ of a system turns out to be inversely proportional to its depth with a constant coefficient, retaining its universality in the anisotropic phase.

Our construction highlights the role of hyperscaling violation in extending the known analytic Yang–Mills solutions to lower bulk dimensions. By appropriately adjusting the hyperscaling-violation exponent, the change in the radial structure of the Yang–Mills equations induced by reducing the bulk dimension can be compensated. The resulting equations retain the same form for $(D, \alpha) = (5, 0), (4, 1/2)$, and $(3, 2)$, admitting the same leading-order condensate profile despite their different asymptotic geometries. Within this analytic family, hyperscaling violation thus provides a direct link between the previously known five-dimensional solution and its lower-dimensional counterparts.

As a final remark,  we explain why we are staying in the small $\delta$ regime and its implication briefly.  The reason why we take $\delta$ to be parametrically small is to obtain the analytic form of the back reaction.  So far,  the only way to get an analytic solution of the back reaction is to perform perturbative expansion in small $\delta$ and $\epsilon$.  This is due to technical reason.  In \cite{Cai:2012nm},  the authors discuss that if $\delta$ is larger than a certain critical value,  it shows first order phase transition whereas if it is smaller than the value,  it represents second order phase transition.
The anisotropic phase transitions in our case are second-order since we chose $\delta$ to be small to obtain the solutions with power expansion.

\section{Holographic model for anisotropic superfluids}
Our holographic model for anisotropic holographic superfluids is based on the action for the Einstein-Scalar-U(1)$\times$SU(2) gauge theory, which is given by
\begin{align}
    S = \frac{1}{\kappa_D^2} \int d^Dx \sqrt{-g} \left( R - \frac{1}{2}g^{\mu\nu}\partial_\mu\phi \partial_\nu\phi + \frac{V_0}{L^2} e^{\gamma\phi} - \frac{\kappa_D^2}{4g_U^2} e^{\lambda_U\phi} F_{\mu\nu}F^{\mu\nu} - \frac{\kappa_D^2}{4g_{YM}^2} e^{\lambda_{YM}\phi} G^a_{\mu\nu}G^{a\,\mu\nu} \right),
\end{align}
where $\kappa_D$ is the $D$-dimensional gravity constant, $x^\mu$ the coordinate variables, with indices $\mu$, $\nu$ running from 0 to $D-1$, and $R$ the curvature scalar.
$\phi$ is a scalar field.
$F_{\mu\nu}$ and $G_{\mu\nu}^a$ are the field strengths of a $U(1)$-gauge field $A_\mu$ and $SU(2)$ gauge fields $B^a_\mu$, respectively.
They are defined by
\begin{align}
	F_{\mu\nu}
		&= \partial_{\mu} A_\nu - \partial_{\nu} A_\mu,
	\\
    G^a_{\mu\nu}
    	&= \partial_\mu B^a_\nu
    		- \partial_\nu B^a_\mu
    		- \epsilon^{abc} B^b_\mu B^c_\nu.
\end{align}
There are also many of theory constants.
We take the length scale $L=1$.
$g_U$ and $g_{YM}$ are gauge couplings.
The constants $V_0$, $\gamma$, $\lambda_U$ and $\lambda_{YM}$ are to be fixed to satisfy equations of motion.
The Einstein equations are denoted by $\mathcal W_{\mu\nu}=0$, the scalar equation by $\mathcal X=0$, the $U(1)$-gauge field equations by $\mathcal Y^\nu=0$, and the $SU(2)$ Yang-Mills field equations by $\mathcal Y^{a\nu}=0$.

\begin{align}
\label{EinsteinEquation}
\begin{split}
     \mathcal{W}_{\mu\nu}
     	&= R_{\mu\nu}
     		- \frac{1}{2}\partial_\mu\phi \partial_{\nu} \phi
     		+ \frac{1}{D-2}\frac{V_0}{L^2} g_{\mu\nu} e^{\gamma\phi}
     		- \frac{\kappa_D^2}{2g_U^2} e^{\lambda_U\phi}
     			\left(
     				F_{\mu\sigma} F_\nu^{\hphantom{\nu}\sigma}
     				- \frac{1}{2D-4}g_{\mu\nu} F_{\alpha\beta}
     					F^{\alpha\beta}
     			\right)
    		\\ &\hphantom{=}
    		- \frac{\kappa_D^2}{2g_{YM}^2} e^{\lambda_{YM}\phi}
    				\left(
    					G^a_{\mu\sigma}G_\nu^{a\hphantom{\nu}\sigma}
    					- \frac{1}{2D-4}g_{\mu\nu} G^a_{\alpha\beta}
    						G^{a\,\alpha\beta}
    				\right),
\end{split}
\end{align}
\begin{align}
\label{ScalarEquation}
     \mathcal{X}
    	= \frac{1}{\sqrt{-g}} \partial_\mu
    			\left(
    				\sqrt{-g}g^{\mu\nu}\partial_\nu\phi
    			\right)
    		+ \frac{V_0}{L^2}\gamma e^{\gamma\phi}
    		- \frac{\kappa_D^2}{4g_U^2}\lambda_U
    			e^{\lambda_U\phi} F_{\mu\nu}F^{\mu\nu}
    		- \frac{\kappa_D^2}{4g_{YM}^2}\lambda_{YM}
    			e^{\lambda_{YM}\phi} G^a_{\mu\nu}G^{a\,\mu\nu},
\end{align}
\begin{align}
    \mathcal{Y}^\nu
    	= \frac{1}{\sqrt{-g}} \partial_\mu
    			\left(
    				\sqrt{-g} e^{\lambda_U\phi} F^{\mu\nu}
    			\right),
\end{align}
\begin{align}
    \mathcal{Y}^{a\nu}
    	= \frac{1}{\sqrt{-g}} \partial_\mu
    			\left(
    				\sqrt{-g} e^{\lambda_{YM}\phi} G^{a\,\mu\nu}
    			\right)
    		+ \epsilon^{abc}e^{\lambda_{YM}\phi}G^{b\,\mu\nu}B^c_\mu.
\end{align}
We adopt the following ansatz for the metric:
\begin{align}
    ds^2
    	&= r^{2\alpha}
    		\left(
	   			- r^{2z} \sigma^2(r)N(r) d\tilde{t}^2
    			+ \frac{dr^2}{r^2 N(r)}
    			+ r^2 \sum_{i,j=1}^d \gamma_{ij}(r)
    				d\tilde{x}^i d\tilde{x}^j
    		\right)
    	\\
    	\label{metric_ansatz}
    	&= r_0^{2\alpha}u^{2\alpha}
    		\left(
    			- u^{2z} N(u) dt^2
    			+ \frac{du^2}{u^2 N(u)}
    			+ u^2 \sum_{i,j=1}^d \gamma_{ij}(u) d{x^i}dx^j
    		\right),
\end{align}
together with
\begin{align}
\label{field_ansatz}
\begin{split}
	B^a\tau^a
		&= b(u) \tau^3 dt
			+ \omega(u) \tau^1 dx_1,
	\\
	A &= a(u) dt,
	\\
	\phi &= \phi(u),
\end{split}
\end{align}
where $u$ is the rescaled holographic coordinate and $1 \leq u < \infty$.  
The metric ansatz is chosen to describe static, translationally
invariant black branes respecting asymptotics with hyperscaling violating and Lifshitz symmetry.  Near the critical point,  the vector condensate selects the $x$-direction(It can be any spatial direction and without loss of generality, one can choose $x$-direction for it).  
Therefore, we allow the metric components parallel and perpendicular to the
condensate to vary independently,  when phase transition occurs.  This is encoded in $\gamma_{ij}(u)$.  The metric function $N(u)$
encodes the horizon, while $\sigma(u)$ allows for an independent
backreaction in the temporal component. The remaining freedom
to redefine the radial coordinate is used to fix a convenient
radial gauge.
Moreover,  the gauge fields and scalar field are chosen in such a way that $\phi(u)$ and $a(u)$ are designed to give asymptotically heperscaling violation and Lifshitz geometry.  $b(u)$ is responsible for chemical potential of black brane and $\omega(u)$ is the vector order near the critical point.  For the solutions considered here, we specialize to $z=1$,
for which the $U(1)$ field strength will vanish.

We note that  
in \eqref{metric_ansatz},  the coordinates are rescaled by
\begin{align}
	u = \frac{r}{r_0},
	\quad
	t = r_0^z\, \tilde{t},
	\quad
	x^i = r_0\, \tilde{x}^i,
\end{align}
which means that $u=1$ is the blackbrane horizon and $u=\infty$ is the holographic boundary.
The simplest solution of the equations of motion is a charged black brane solution with hyperscaling violation factor $\alpha$ and Lifshitz scaling factor $z$, which is spatially isotropic.
The metric is given by
\begin{equation}
\label{metric_components_without_backreaction}
   \sigma(u)=1,
   \,
   N(u) = 1 - \frac{m r_0^{-(d\alpha + z + d)}}{u^{d\alpha + z + d}}
   		+\frac{\delta^2\mu^2}{u^{2(d\alpha + z + d-1)}},
   \,
   \textrm{and\ }
   \gamma_{ij}=\delta_{ij},
\end{equation}
where $d=D-2$ is the spatial dimensions, $m$ is the black brane mass, $\mu$ is the chemical potential, $\delta=\frac{\kappa_D}{g_{YM}}r_0$,  and the horizon of the black brane is located at $u=1$.
The mass parameter, $m$ of the black brane is proportional to the
energy density $\langle T_{tt}\rangle$ of the dual boundary theory,
with a proportionality factor involving the $D$-dimensional
gravitational constant\cite{Balasubramanian:1999re}.
The chemical potential $\mu$ is a $SU(2)$ gauge potential defined by
\begin{align}
	\mu	= b ( \infty ) - b ( 1 ),
\end{align}
where $b(u)$ is the temporal component of $B^3$ as given in \eqref{field_ansatz}.
The mass $m$ is determined by requiring the metric component $N$ to be zero at the horizon, $u=1$.
It gives
\begin{align}
\label{black_brane_mass_from_N_1}
	1 - \frac{m}{r_0^{d\alpha + z + d}} + \delta^2\mu^2
	=
		0,
\end{align}
so
\begin{align}
\label{black_brane_mass_from_N_2}
	m
	=
		r_0^{d\alpha + z + d} \left( 1 + \delta^2\mu^2 \right).
\end{align}
Therefore,
\begin{align}
\label{N0-background-solution}
	N(u)
	=
		1 - \frac{1}{u^{d\alpha + z + d}}
   	+
   		\frac{\delta^2\mu^2}{u^{d\alpha + z + d}}
   		\left( 1 - \frac{1}{u^{d\alpha+z+d-2}} \right).
\end{align}
The solutions for the scalar and gauge fields are given by
\begin{align}
         \label{phi-u-solution-BG}
	\phi(u) &= \phi_0 + \sqrt{2d(\alpha+1)(\alpha+z-1)}\log u,
	\\
	\label{b-u-solution}
	b(u)
		&=
		\mu r_0^\alpha \sqrt{\frac{2d(\alpha+1)}{d\alpha+z+d-2}}
		e^{\sqrt{\frac{\alpha+z-1}{2d(\alpha+1)}}\phi_0}
		\left( 1 - u^{-d\alpha-z-d+2} \right),
	\\
	G^1_{ux_1} &= 0,
	\\
	G^3_{ut}
		&= \mu r_0^\alpha \sqrt{2d(\alpha+1)(d\alpha+z+d-2)}
		e^{\sqrt{\frac{\alpha+z-1}{2d(\alpha+1)}}\phi_0}
		u^{-d\alpha-z-d+1},
	\\
	F_{ut}
		&= \frac{g_U}{\kappa_D} r_0^\alpha
			\sqrt{2(z-1)(z+d+d\alpha)}
			e^{
				\frac{ -\alpha(1-d) + d }
					{ \sqrt{2d(1+\alpha)(\alpha+z-1)} }
				\phi_0
			}
			u^{d\alpha+d+z-1}.
\end{align}
The theory constants should be fixed as
\begin{align}
\begin{split}
    \lambda_U
    	&= - \frac{2\alpha (d - 1) + 2d}
    		{\sqrt{2d (\alpha + 1)(\alpha + z - 1)}},
    \\
    \lambda_{YM}
    	&= \sqrt{\frac{2(\alpha + z - 1)}{d(\alpha + 1)}},
    \\
    \gamma
    	&= -\frac{2\alpha}{\sqrt{2d (\alpha + 1)(\alpha + z - 1)}},
    \\
    V_0
    	&= r_0^{-2\alpha}e^{-\gamma \phi_0}(d\alpha+z+d-1)(d\alpha+z+d)
\end{split}
\end{align}

\section{Anisotropic solution in the rescaled coordinate frame}
In this section, we want to get analytic solutions which show spatial anisotropy.
The solution will be obtained by employing perturbative method by assuming that the back reactions from the $SU(2)$-Yang-Mills fields is small.
We take a limit that the spatial anisotropy $\omega(u)$ is small but finite, which is suppressed by a dimensionless small parameter $\epsilon$.
Also, $\delta = \frac{\kappa_D}{g_{YM}}r_0$ is taken to be very small.
As explained in the introduction, this limit enables us to expand the solutions in powers of $\delta$, but also restricts the phase transition to be second-order. 

Let us consider the Yang-Mills equations first.
The non-zero components of the field-strength of the Yang-Mills fields are given by
\begin{alignat}{4}
    G^3_{ut}
    	&= \partial_u B_t^3 - \partial_t B_u^3
    		- \epsilon^{312} B_u^1 B_t^2
    		- \epsilon^{321} B_u^2 B_t^1
    	&&= \partial_u b(u),
    \\
    G^1_{u{x_1}}
    	&= \partial_u B_{x_1}^1 - \partial_{x_1} B_u^1
    		- \epsilon^{123} B_u^2 B_{x_1}^3
    		- \epsilon^{132} B_u^3 B_{x_1}^2
    	&&= \partial_u w(u),
    \\
    G^2_{t{x_1}}
    	&= -\epsilon^{213} B_t^1 B_{x_1}^3
    		- \epsilon^{231} B_t^3 B_{x_1}^1
    	= -b(u) w(u).
\end{alignat}
%
The equations of motion are written as
\begin{align}
    e^{-\lambda_{YM}\phi_0}\mathcal{Y}^{3t}
    	&= u^{-\alpha(d + 2) - z - d + 1} \partial_u
    			\left[
    				u^{d(\alpha + 1) + z - 1} \partial_u b(u)
    			\right]
    		- \frac{u^{-2(\alpha + 2)}}{1 - u^{-d\alpha - z - d}}
    			b(u) w^2(u),
    \\
    e^{-\lambda_{YM}\phi_0}\mathcal{Y}^{1{x_1}}
    	&= u^{-\alpha(d + 2) - z - d + 1} \partial_u
    		\left[
    			u^{d\alpha + d + 3z - 3}
    			\left( 1 - u^{-d\alpha - z - d} \right)
    			\partial_u w(u)
    		\right]
    		+ \frac{u^{-2\alpha - 4}}{1 - u^{-d\alpha - z - d}} b^2(u) w(u)
\end{align}
When $\omega(u)=0$, the solution for $b(u)$ is given by (\ref{b-u-solution}).
Again, this solution is spatially isotropic.
Now, let us turn on $\omega(u)$ to break such an isotropy.
We solve the Yang-Mills equations with power expansion order by order, in a small and dimensionless parameter $\epsilon$.
More precisley, we take 
\begin{align}
\label{perturbative_wb}
    w(u) = \sum_{n=0}^\infty \epsilon^{2n+1} w_{2n+1}(u)
    \qquad \textrm{and} \qquad
    b(u) = \sum_{n=0}^\infty \epsilon^{2n} b_{2n}(u),
\end{align}

\subsection{A new class of the analytic solution with spatial anisotropy}

There are special conditions under which solutions for $b(u)$ and $w(u)$ can be obtained analytically.
To the best of our knowledge, the other conditions leading to analytic solutions have not yet identified.
Therefore, while the following three conditions are sufficient, their necessity remains an open issue.
The conditions are that
\begin{itemize}
\item condition 1: $d (\alpha + 1 )= 3$,
\item condition 2: $z=1$,
\item condition 3: $
\mu r^\alpha_0\sqrt{3}e^{-\sqrt{\frac{\alpha}{6}}\phi_0}=4$

\end{itemize}
We note that the condition 1 and 2 do not violate the null energy condition\cite{Alishahiha:2012qu},
\begin{align}
    T_{\alpha\beta}\eta^\alpha \eta^{\beta}\sim(\alpha + 1)(\alpha + z - 1) \ge 0 {\ \ }\rightarrow{\ \ }
    \frac{3}{d}\left(\frac{3}{d} - 1\right) \ge 0,
\end{align}
only when $d$ is 1, 2, or 3. $T_{\alpha\beta}$ is stress energy tensor of the spacetime and $\eta^\alpha=(\sqrt{g^{uu}},\sqrt{g^{tt}},\vec 0)$ is a null vector.
In this case, $\mathcal{Y}^{3t}$ and $\mathcal{Y}^{1x}$ equations can be written as
\begin{alignat}{2}
\label{Yang-Mills-y3t}
    \mathcal{Y}^{3t}
    &=0
    = u^{-2(\alpha + 2)} \left[ u\, \partial_u \left( u^3 \partial_u b(u) \right) - \frac{b(u) w^2(u)}{1 - u^{-4}} \right],
    \\
\label{Yang-Mills-y1x}
    \mathcal{Y}^{1x}
    &=0
    = u^{-2(\alpha + 2)} \left\{ u\, \partial_u \left[
    u^{3} (1 - u^{-4}) \partial_u w(u) \right]
    + \frac{b^2(u) w(u)}{1 - u^{-4}} \right\}.
\end{alignat}
Solving the set of equations (\ref{Yang-Mills-y3t}) and (\ref{Yang-Mills-y1x}) is Sturm–Liouville problem since they have solutions for a certain value of $\mu$.
The condition 3 is required for the solutions to exist.
We choose
\begin{equation}
         \label{mu-phi0-condition-BG}
	\mu = \frac{4}{\sqrt{3}}
	\qquad
	\textrm{and}
	\qquad
	\phi_0 = \sqrt{6\alpha} \log r_0.
\end{equation}
Now, we list the perturbative solutions, which are given by
\begin{align}
\begin{split}
    b_0(u)
    &=
    	4 \left( 1 - u^{-2} \right),
    \\
    b_2(u)
    &=
    	\frac{71}{6720} \left( 1 - \frac{1}{u^2} \right)
    	+ \frac{5 + 7u^2 - 9u^4 - 3u^6 }{96u^2(1+u^2)^3},
\end{split}
\\
    w_1(u)
    &=
    	\frac{u^2}{ ( 1 + u^2 )^2 },
    {\rm \ \ and\ so\ on.}
\end{align}

Before closing this subsection, we note that solutions are not restricted to $\mu=4/\sqrt{3}$. According to the Sturm–Liouville analysis of the Yang-Mills equations in \cite{Park:2016wch}, for $(D,\alpha,z)=(5,0,1)$, 
the lowest eigenvalue of $\sqrt{3}\mu$ at the critical point is bounded above by $4.09206$, which is close to $4$. 
The cases for $(D,\alpha,z)=(4,1/2,1)$, and $(3,2,1)$,  the same results will come out since they share the same form of the Yang-Mills equations.
Although the authors did not explore the full space of trial functions, their results suggest that the lowest eigenvalue may be $\sqrt{3}\mu=4$. This, however, does not rule out additional solutions with $\sqrt{3}\mu>4$. Such solutions may be obtainable analytically or numerically.
\subsection{Remarks on Yang-Mills solutions}
When the thermodynamic phase transition between normal-fluid and super-fluid phases occurs,  the vector order $\omega(u)$ appears. This vector order breaks global symmetry of spatial rotations and inversions. Without loss of generality, one can set the direction of the vector order is along $x_1$-direction. This breaks $O(d)$ symmetry to $O(d-1)$, when $d>2$. Again, $d$ is the number of spatial dimensions. When $d=2$, $O(2)$ symmetry is broken down to $Z(2)$, due to the order parameter. When $d=1$, we have only one spatial dimension enjoying parity symmetry as $x_1\rightarrow-x_1$, namely $Z(2)$ symmetry. This is also broken completely with non-zero vector order, $\omega(u)$ in anisotropic phase.

Our analytic solutions share $z=1$ but the values of $\alpha$ are different from one another. When $d=3$, the dual gravity spacetime is asymptotically AdS$_5$. However, $d=2$ and $d=1$ cases are asymptotically hyperscaling violation geometries with $\alpha=1/2$ and $\alpha=2$ respectively. The results are summerized in the Table \ref{symmetry-mech-table}.

 We must note that the analytic solution for Yang-Mills field and for leading order back reactioins to background spacetime which we will discuss in the next section in $D=d+2=5$ is already obtained in \cite{Basu:2011tt}. However, the analytic solutions being asymptotically heperscaling violation geometry in $D=d+2=4$ and $D=d+2=3$ are newly obtained in this paper.

\setlength\heavyrulewidth{0.25ex}
\setlength{\tabcolsep}{8pt}
\newcolumntype{?}{!{\vrule width 1.75 pt}}
\begin{table}[H]
\centering
\renewcommand{\arraystretch}{2}
{\normalsize
\begin{tabular}{|c|c|c|c|c|}
\noalign{\hrule height 2 pt}
	\multicolumn{1}{c|}{$\boldsymbol{D=d+2}$} &
	\multicolumn{1}{c|}{$\boldsymbol{\mu}$} &
	\multicolumn{1}{c|}{$\boldsymbol{\alpha}$} &
	\multicolumn{1}{c|}{$\boldsymbol{z}$} &
	\multicolumn{1}{c}{\textbf{Symmetry Breaking}} \\
\noalign{\hrule height 2 pt}
	\multicolumn{1}{c|}{5} &
	\multicolumn{1}{c|}{$4/\sqrt{3}$} &
	\multicolumn{1}{c|}{0} &
	\multicolumn{1}{c|}{1} &
	\multicolumn{1}{c}{${O}(3) \rightarrow {O}(2)$} \\
\noalign{\hrule height 1 pt}
	\multicolumn{1}{c|}{4} &
	\multicolumn{1}{c|}{$4/\sqrt{3}$} &
	\multicolumn{1}{c|}{$1/2$} &
	\multicolumn{1}{c|}{1} &
	\multicolumn{1}{c}{${O}(2) \rightarrow {Z}(2)$} \\
\noalign{\hrule height 1 pt}
	\multicolumn{1}{c|}{3} &
	\multicolumn{1}{c|}{$4/\sqrt{3}$} &
	\multicolumn{1}{c|}{2} &
	\multicolumn{1}{c|}{1} &
	\multicolumn{1}{c}{${Z}(2)$ sym broken} \\
\noalign{\hrule height 2 pt}
\end{tabular}
}
\caption{
	Symmetry breaking mechanism.
}
\label{symmetry-mech-table}
\end{table}

\section{Anisotropic geometry, asymptotically hyperscaling violation at $D=3$ and $D=4$}
In this section, we obtain the leading-order backreactions induced by the vector order $\omega(u)$, near the critical point.
They give a new class of the solutions, which are asymptotically hyperscaling violation geometry in $D$-dimensional spacetime.
The analytic solutions are obtained for the cases $D=d+2=3$ and $D=d+2=4$.

\subsection{The solution for $\alpha=2$ and $z=1$ at $D=3$ case}
As we discussed in the previous section, when $D=d+2=3$, there exists an analytic solution if $\alpha=2$ and $z=1$.
In this case, we have
\begin{align}
    \lambda_U = - \frac{1}{\sqrt 3},
    \quad
    \lambda_{YM} = \frac{2}{\sqrt 3},
    \quad
    \gamma = -\frac{2}{\sqrt 3},
    \quad
    \textrm{and}
    \quad
    V_0=12.
\end{align}
To get the the analytic solutions, we rewrite the metric as
\begin{align}
\label{metric_form_with_backreaction}
	ds^2 =
		- r_0^4 u^6 N(u)\, \sigma^2(u) dt^2
		+
		\frac{r_0^4 u^2}{N(u)} du^2
		+
		r_0^4 u^6 H^2(u) dx^2,
\end{align}
where the functions $N,\sigma,H$ and the scalar field $\phi$ are defined up to $\epsilon^2\delta^2$-order by
\begin{align}
\label{d1solutionsform}
\begin{split}
    N(u) &=
    	1
    	+
    	\frac{16\delta^2}{3u^6}
    	-
    	\frac{3+16\delta^2}{3u^4}
    	+
    	\epsilon^2\delta^2N_{2}(u),
    \\
    H(u) &=
    	1
    	+
    	\epsilon^2\delta^2H_{2}(u),
    \\
    \sigma(u) &=
    	1
    	+
    	\epsilon^2\delta^2\sigma_{2}(u)
    \\
    \phi(u) &=
    	2\sqrt{3}\log r_0
    	+
    	2\sqrt{3}\log u
    	+
    	\epsilon^2\delta^2\phi_{2}(u).
\end{split}
\end{align}
together with
\begin{equation}
	F_{ut}=0.
\end{equation}
We note that these solutions are perturbative, which capture the leading-order backreactions from the vector order $\omega(u)$.
Also, we expect that one of the functions appearing at order $\epsilon^2\delta^2$ can be fixed arbitrarily, because there is a gauge freedom in gravity theories.
Actually, we obtained the solutions for $N_2, \sigma_2$, and $\phi_2$ in terms of $H_2$ and it turned out that $H_2$ can be chosen freely.
Therefore, we carry out the following calculations with $H_2(u)$ set to zero from the outset.

To begin with, we rewrite the Einstein equations $\mathcal{W}_{tt}, \mathcal{W}_{uu}$ and $\mathcal{W}_{xx}$ given in \eqref{EinsteinEquation} into more convenient forms.
With the newly defined metric, they can be expressed in terms of the backreaction terms:

\begin{align}
\begin{split}
    \mathcal{E}_1
    	&\equiv \frac{2u^5}{-1+u^4} \mathcal{W}_{tt}
    	\\
    	&=
    		- \frac{4u(-1+u^2)^3}{(1+u^2)^5}
    		+ 24 u^3 N_2(u)
    	   	+ 11 u^4 N_2'(u)
    		+ u^5 N_2''(u)
    		\\ &\hphantom{=-}
    		+ 4 (-1 + 4 u^4) \sigma_2'(u)
    		+ 2u (-1 + u^4) \sigma_2''(u)
    		+ 16\sqrt{3} u^3 \phi_2(u),
\end{split}
\\
\begin{split}
    \mathcal{E}_2
    	&\equiv 2u(-1+u^4) \mathcal{W}_{uu}
		\\
		&=
    		- \frac{16 u^3 (-1 + u^2)}{(1 + u^2)^5}
    		- 24 u^3 N_2(u)
    		- 11 u^4 N_2'(u)
    		- u^5 N_2''(u)
    	\\ &\hphantom{=-}
    		- 2 (1 + 5 u^4) \sigma_2'(u)
    		- 2 u (-1 + u^4) \sigma_2''(u)
    		- 16\sqrt{3} u^3 \phi_2(u)
    		- 4\sqrt{3} (-1 + u^4) \phi_2'(u),
\end{split}
\\
\begin{split}
    \mathcal{E}_3
    	&\equiv u\, \mathcal{W}_{xx}
    	\\
    	&=
    		\frac{279+1116u^2+1674u^4-1124u^6-281u^8}{420u^3(1+u^2)^4}
    		- 12u^3 N_2(u)-3u^4N_2'(u)
    		\\ &\hphantom{=-}
    		- 3(-1+u^4) \sigma_2'(u)
    		- 8\sqrt{3}u^3\phi_2(u).
\end{split}
\end{align}
The scalar equation $\mathcal{X}$ given in \eqref{ScalarEquation} is rewritten into
\begin{align}
\begin{split}
    \mathcal{E}_S
    	&\equiv \frac{\sqrt{3}}{2} r_0^4 u^7 \mathcal{X}
    	\\
    	&=
    		- \frac{279+1395u^2+1950u^4+6430u^6-7285u^8+559u^{10}}
    				{420u^3(1+u^2)^5}
    		+ 12u^3N_2(u) + 3u^4N_2'(u)
   		\\ &\hphantom{=-}
   		+ 3 (-1+u^4) \sigma_2'(u)
   		+ 8\sqrt{3}u^3\phi_2(u)
   		+ \frac{\sqrt{3}}{2} (-1+5u^4)\phi_2'(u)
   		+ \frac{\sqrt{3}}{2} u (-1+u^4)\phi_2''(u).
\end{split}
\end{align}
First of all, the differential equation for $\phi_2(u)$ can be obtained as follows.
\begin{align}
\begin{split}
	\mathcal{E}_3 + \mathcal{E}_S
	&=
	\frac{2u(1-7u^2+7u^4-u^6)}{(1+u^2)^5}
	+
	\frac{\sqrt{3}}{2} (-1+5u^4)\phi_2'(u)
	+
	\frac{\sqrt{3}}{2} u(-1+u^4)\phi_2''(u)
	= 0.
\end{split}
\end{align}
We integrate both sides twice and $\phi_2(u)$ is given by
\begin{equation}
    \phi_2(u)
    =
    \frac{-1 + 2u^2}{ 6\sqrt{3}(1+u^2)^4 }
    +
    \frac{C_\phi}{2\sqrt{3}} \log \left( 1 - \frac{1}{u^4} \right)
    +
    D_\phi,
\end{equation}
where $C_\phi$ and $D_\phi$ are the constants of the first and second integrations, respectively.

The other two equations $\mathcal{E}_1$ and $\mathcal{E}_2$ give the solution for $\sigma_2(u)$.
We integrate both sides of
\begin{align}
	\frac{\mathcal{E}_1 + \mathcal{E}_2}{6(-1+u^4)}
	=
	\frac{4\, C_\phi}{3u(1-u^4)}
	+
	\frac{2u(-3+u^2)}{3(1+u^2)^5}
	+
	\sigma_2'(u)
	=
	0.
\end{align}
It gives
\begin{align}
    \sigma_2(u)
    &=
    \frac{-2+u^2}{9(1+u^2)^4}
    +
    \frac{C_\phi}{3}\log\left(1-\frac{1}{u^4}\right)
    +
    C_\sigma
    ,
\end{align}
where $C_\sigma$ is the constant of integration.

Because we have fixed $H_2(u)=0$, there remains only $N_2(u)$.
By integrating both sides of
\begin{align}
\begin{split}
	\frac{\mathcal{E}_3}{3}
	&=
	- \frac{4C_\phi}{3u}
	-
	\frac{8D_\phi}{\sqrt{3}}u^3
	+
	\frac{93+372u^2+1398u^4-1308u^6-187u^8}{420u^3(1+u^2)^4}
	-
	\frac{4}{3}C_\phi u^3\log\left(1-\frac{1}{u^4}\right)
	\\ &\hphantom{=}
	-4 u^3 N_2(u) - u^4 N_2'(u)
	=
	0,
\end{split}
\end{align}
we obtain
\begin{align}
\begin{split}
    N_2(u)
    &=
    -\frac{2D_\phi}{\sqrt{3}}
    +
    \frac{C_N}{u^4}
    +
    \frac{-279-557u^2+2523u^4+561u^6}{2520u^6(1+u^2)^3}
    -\frac{C_\phi}{3}\left(1-\frac{1}{u^4}\right)\log\left(1-\frac{1}{u^4}\right)
\end{split}
\end{align}
where $C_N$ is the constant of integration.

The values of the constants are determined by the following conditions.
First of all, these functions are required to be regular at the horizon, $u=1$.
Hence we choose $C_\phi=0$ to eliminate the divergent logarithmic terms.
Secondly, the spacetime is required to be asymptotically hyperscaling violation geometry.
Therefore $C_\sigma = D_\phi = 0$, because the $\epsilon^2\delta^2$-order terms should vanish at $u\rightarrow\infty$.
Lastly, we fix the location of the horizon at $u=1$.
We need $N_2(u=1)=0$ for this condition and it is satisfied when $C_N=-281/2520$.
Finally we have
\begin{align}
    N(u) &=
    1 + \frac{16\delta^2}{3u^6} - \frac{3+16\delta^2}{3u^4}
    +\epsilon^2\delta^2
    	\frac{-279 - 838 u^2+ 1680 u^4 - 282 u^6 - 281 u^8}
    		{ 2520 u^6 \left( 1+ u^2 \right)^3 },
    \\
    H(u) &= 1,
    \\
    \sigma(u) &=
    	1
    	+
    	\epsilon^2\delta^2\, \frac{ -2 + u^2 }{ 9 ( 1 + u^2 )^4 },
    \\
    \phi(u) &=
    	2\sqrt{3}\log r_0 + 2\sqrt{3}\log u
    	+ \epsilon^2\delta^2\, \frac{ -1 + 2u^2 }{ 6\sqrt{3} (1+u^2)^4 }.
\end{align}
With these solutions,  we compute the holographic entanglement entropy of a line-shaped subsystem on the boundary in the next section.



\subsection{The solution for $\alpha=\frac{1}{2}$ and $z=1$ at $D=4$ case}
When $D=d+2=4$, an analytic solution exists if $\alpha=1/2$ and $z=1$. In this case, we have
\begin{align}
    \lambda_U = - \frac{5}{\sqrt 3},
    \quad
    \lambda_{YM} = \frac{1}{\sqrt 3},
    \quad
    \gamma = - \frac{1}{\sqrt 3},
    \quad
    \textrm{and}
    \quad
    V_0=12,
\end{align}
together with $F_{ut}=0$.
We follow a similar approach as in the previous section.
Since there is an additional spatial dimension, we introduce another function
\begin{equation}
	J(u) = 1 + \epsilon^2\delta^2 J_{2}(u)
\end{equation}
such that the metric is rewritten as
\begin{align}
	ds^2
	=
		-r_0 u^3 N(u) \sigma^2(u) dt^2
		+
		\frac{r_0}{ u N(u) } du^2
		+
		r_0u^3
			\left(
				H^2(u) dx^2	+ J^2(u) dy^2
	 		\right).
\end{align}
The other functions $N(u), \sigma (u), H(u)$ and $\phi (u)$ have the same forms as given in \eqref{d1solutionsform}, up to $\epsilon^2\delta^2$-order.
As in the previous subsection, it turns out that $H_2$ is free to choose, hence we set $H_2(u) = -J_2(u)$.
The Einstein equations $\mathcal{W}_{tt}, \mathcal{W}_{uu}, \mathcal{W}_{xx}$ and $\mathcal{W}_{yy}$ in \eqref{EinsteinEquation} are rewritten as follows, in terms of backreactions.
\begin{align}
\begin{split}
	\mathcal{E}_1
		&\equiv 840u^8(1+u^2)^3\mathcal{W}_{tt}		
		\\
		&=
			- 279 - 837 u^2 - 1398 u^4 + 3638 u^6 - 3 u^8 - 1121 u^{10}
		\\ &\hphantom{=-}
			+ 420 u^6 (-1 + u^2) (1+u^2)^4
				\left(
					12 N_2(u) +	8 u N_2'(u)	+ u^2 N_2''(u)
				\right)
		\\ &\hphantom{=-}
			+ 420 u^3 (-1 + u^2) (1+u^2)^4
				\left[
					(-1 + 13 u^4) \sigma_2'(u)
					+ 2 u (-1 + u^4) \sigma_2''(u)
				\right]
		\\ &\hphantom{=-}
			+ 1680\sqrt{3} u^6 (-1 + u^2) (1 + u^2)^4 \phi_2(u),
\end{split}
\\
\begin{split}
	\mathcal{E}_2
		&\equiv 840u^4(-1+u^2)(1+u^2)^5 \mathcal{W}_{uu}
		\\
		&=
			- 279 - 1116 u^2 - 834 u^4 + 1124 u^6 - 559 u^8
			\\ &\hphantom{=-}
			- 420 u^6 (1 + u^2)^4
				\left(
					12 N_2(u) + 8 u N_2'(u)	+ u^2 N_2''(u)
				\right)
			\\ &\hphantom{=-}
			- 420 u^3 (1 + u^2)^4
				\left[
					(5 + 7 u^4) \sigma_2'(u)
					+ 2 u (-1 + u^4) \sigma_2''(u)
				\right]
		\\ &\hphantom{=-}
			- 840\sqrt{3} u^3 (1 + u^2)^4
				\left[
					2 u^3 \phi_2(u) + (-1 + u^4) \phi_2'(u)
				\right],
\end{split}
\\
\begin{split}
	\mathcal{E}_3
		&\equiv 840 u^4 (1+u^2)^5 \mathcal{W}_{xx}
		\\
		&=
			279 + 1395 u^2 + 3630 u^4 - 5330 u^6
				+ 4475 u^8 - 1121 u^{10}
			\\ &\hphantom{=-}
			- 1260 u^6 (1 + u^2)^5 ( 4 N_2(u) + u N_2'(u) )
			- 1260 u^3 (-1 + u^2) (1 + u^2)^6 \sigma_2'(u)
			\\ &\hphantom{=-}
			+ 840 u^3(1+u^2)^5(-1+5u^4) J_2'(u)
			+ 840 u^4(1+u^2)^5(-1+u^4)^6 J_2''(u)
			\\ &\hphantom{=-}
			- 1680\sqrt{3} u^6 (1 + u^2)^5 \phi_2(u),
\end{split}
\\
\begin{split}
	\mathcal{E}_4
		&\equiv 840 u^4 (1+u^2)^5 \mathcal{W}_{yy}
		\\
		&=
			279 + 1395 u^2 + 1950 u^4 + 6430 u^6
				- 7285 u^8 + 559 u^{10}
			\\ &\hphantom{=-}
			- 1260 u^6 (1 + u^2)^5 ( 4 N_2(u) + u N_2'(u) )
			- 1260 u^3 (-1 + u^2) (1 + u^2)^6 \sigma_2'(u)
			\\ &\hphantom{=-}
			- 840 u^3 (1+u^2)^5(-1+5u^4) J_2'(u)
			- 840 u^4 (1 + u^2)^5 (-1 + u^4) J_2''(u)
			\\ &\hphantom{=-}
			- 1680\sqrt{3} u^6 (1 + u^2)^5 \phi_2(u).
\end{split}
\end{align}
\allowdisplaybreaks[0]
The scalar equation \eqref{ScalarEquation} is rewritten into
\begin{align}
\begin{split}
	\mathcal{E}_S
		&\equiv -420\sqrt{3}r_0u^7(1+u^2)^5 \mathcal{X}
		\\
		&=
		279 + 1395 u^2 + 1950 u^4 + 6430 u^6 - 7285 u^8 + 559 u^{10}
		\\ &\hphantom{=}
		-
		1260 u^6 (1 + u^2)^5 (4 N_2(u) + u N_2'(u))
		-
		1260 u^3 (-1 + u^2) (1 + u^2)^6 \sigma_2'(u)
		\\ &\hphantom{=}
		-
		420\sqrt{3} u^3 (1 + u^2)^5
			\left[
				4 u^3 \phi_2(u)
				+
				(-1 + 5u^4) \phi_2'(u)
				+
				u (-1 + u^4) \phi_2''(u)
			\right].
\end{split}
\end{align}
First of all, the differential equation for $J_2(u)$ is given by
\begin{align}
	\frac{\mathcal{E}_3 - \mathcal{E}_4}{1680 u^3 (1 + u^2)^5}
	=
		\frac{u-7u^3+7u^5-u^7}{(1+u^2)^5}
		+
		(-1+5u^4)J_2'(u)
		+
		u(-1+u^4)J_2''(u)
	=
		0.
\end{align}
We integrate both sides twice and obtain
\begin{align}
	J_2(u)
	=
		\frac{-1+2 u^2}{24 (1 + u^2)^4}
	+
		\frac{1}{4} C_J \log \left( 1 - \frac{1}{u^4} \right)
	+
		D_J,
\end{align}
where $C_J$ and $D_J$ are the constants of the first and second integration, respectively.

The solution for $\phi_2(u)$ is obtained from $\mathcal{E}_3$ and $\mathcal{E}_S$.
\begin{align}
	\frac{\mathcal{E}_3 - \mathcal{E}_S}{420\sqrt{3} u^3 (1 + u^2)^5}
	=
	\frac{2(u-7u^3+7u^5-u^7)}{\sqrt{3}(1+u^2)^5}
	+
	(-1+5u^4)\phi_2'(u)
	+
	u(-1+u^4)\phi_2''(u)
	= 0.
\end{align}
By integrating both sides twice, we get
\begin{equation}
\label{phi2-solutionD=4}
	\phi_2(u)
	=
		\frac{-1 + 2 u^2}{12\sqrt{3} (1 + u^2)^4}
	+
		\frac{1}{4} C_\phi \log \left(1 - \frac{1}{u^4}\right)
	+
		D_\phi,
\end{equation}
where $C_\phi$ and $D_\phi$ are the constants of the first and second integration, respectively.

For $\sigma_2(u)$, we integrate both sides of
\begin{align}
	\frac{\mathcal{E}_1 - (1-u^2)\, \mathcal{E}_2}
	{2520 u^3 (-1 + u^2)^2 (1+ u^2)^5}
	=
		\frac{C_\phi}{\sqrt{3}u(1-u^4)}
	-
		\frac{3u+u^3}{3(1+u^2)^5}
	+
		\sigma_2'(u)
	=
		0.
\end{align}
It gives
\begin{align}
\begin{split}
	\sigma_2(u)
	&=
		-
			\frac{5+2u^2}{36 (1 + u^2)^4}
		+
			\frac{1}{4\sqrt{3}} C_\phi
				\log \left( 1 - \frac{1}{u^4} \right)
		+
			C_\sigma,
\end{split}
\end{align}
where $C_\sigma$ is the constant of integration.

The remaining equation is given by
\begin{align}
\begin{split}
	\frac{\mathcal{E}_S}{1260u^3(1+u^2)^5}
	&=
	-
		\frac{C_\phi}{\sqrt{3}u}
	-
		\frac{4u^3D_\phi}{\sqrt{3}}
	-
		\frac{C_\phi u^3}{\sqrt{3}}
			\log \left( 1 - \frac{1}{u^4} \right)
	\\
	&+
		\frac{93+372u^2+978u^4-608u^6-327u^8}{420u^3(1+u^2)^4}
	-
		4 u^3 N_2(u)
	-
		u^4 N_2'(u)
	=
		0.
\end{split}
\end{align}
From this equation we lastly obtain
\begin{align}
\label{N2-solutionD=4}
\begin{split}
	N_2(u) =
		-\frac{D_\phi}{\sqrt{3}}
		+
		\frac{C_N}{u^4}
		-
		\frac{279 + 347 u^2 - 1893 u^4 - 981 u^6}
			{2520 u^6 (1 + u^2)^3}
		-
		\frac{C_\phi}{4\sqrt{3}}\left(1-\frac{1}{u^4}\right)
		\log\left(1-\frac{1}{u^4}\right),
\end{split}
\end{align}
where $C_N$ is constant of integration.

The values of the constants are determined by the same conditions as in the previous subsection.
These functions are regular at the horizon, the spacetime should be asymptotically hyperscaling-violating, and the location of the horizon is fixed at $u=1$.
These conditions give $C_N=-281/2520$, with the other constants being zero.
Finally we have
\begin{align}
	N_2(u)
	&=
		\frac{ -279 -628 u^2 + 1050 u^4 + 138 u^6 - 281 u^8 }
			{ 2520 u^6 ( 1 + u^2 )^3 },
	\\
	\sigma_2(u)
	&=
		- \frac{5+2u^2}{36(1+u^2)^4},
	\\
	J_2(u)
	&=
		- \frac{1-2u^2}{24(1+u^2)^4}
	=
		-H_2(u),
	\\
	\phi_2(u)
	&=
		- \frac{1-2u^2}{12\sqrt{3}(1+u^2)^4}.
\end{align}
With these solutions, we compute the holographic entanglement of a narrow strap-shaped system on the boundary in the next section.

Before closing this subsection, we remark one thing. Since the backreaction is induced by the condensate, we require it to describe a change of state in dual field theory without modifying the prescribed boundary sources. We therefore exclude non-normalizable deformations that would alter the boundary geometry or the scalar/vector source, retaining only the response associated with the condensate. Accordingly, the metric corrections must be subleading relative to the background metric near the boundary, while the scalar/vector correction must preserve the prescribed asymptotic scalar/vector data respectively.

We also need to take into account horizon regularity to ensure that the stress-energy tensor of the perturbation remains finite in the given background.  A divergence would invalidate the perturbative approximation, because the backreaction could no longer be treated as a small correction to the background.  For more general discussion of the boundary conditions,  see for example \cite{Kanitscheider:2008kd,Kanitscheider:2009as}.

To demonstrate this briefly, let us take trace of Einstein equation(\ref{EinsteinEquation}). With this, one can compute curvature scalar $R$ in terms of other fields. Especially, if $D=4$, the contribution from gauge fields disappears and the curvature scalar has a form of
\begin{equation}
R=\frac{1}{2}g^{uu}(\partial_u \phi(u))^2-24e^{-\frac{\phi(u)}{\sqrt{3}}},
\end{equation}
where $g^{uu}=uN(u)/r_0$. Suppose we turn on the terms proportional to $C_\phi$ in the solutions of (\ref{phi2-solutionD=4}) and (\ref{N2-solutionD=4}). Then, the leading backreaction to the curvature scalar $R$ should be logarithmically divergent as\begin{equation}
r^2_0 R=-24+\epsilon^2\delta^2 2\sqrt{3}\log(u-1)+{\rm finite\ terms}
\end{equation}
when $u$ approaches the black brane horizon, $u=1$, where we use the relations, (\ref{N0-background-solution}), (\ref{phi-u-solution-BG}), (\ref{mu-phi0-condition-BG}) to derive the expression of $R$.  Therefore, to remove this divergence, we must take $C_\phi=0$.  

\section{Holographic entanglement entropies}
\subsection{$D=3$ case: A line-shaped subsystem lying along $x$-axis}
	\label{subsection_D3line_area_compute}
\subsubsection{Holographic entanglement entropy}
\begin{figure}[h]
	\label{D3line}
	\centering
	\includegraphics[scale=0.25]{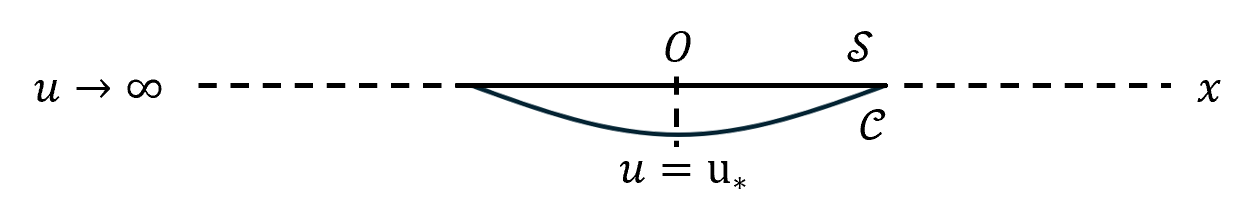}
	\caption{}
\end{figure}
\noindent
Consider a line-shaped subsystem $\mathcal{S}$ on the $u=\infty$ boundary.
It lies along the $x$-axis at $u = \infty$, from $x = - \frac{L}{2}$ to $x = \frac{L}{2}$.
To obtain the holographic entanglement entropy of $\mathcal{S}$, we consider a curve $\mathcal{C}$ that hangs down in the bulk direction from the endpoints of $\mathcal{S}$.
We need to compute the minimal area of $\mathcal{C}$ \cite{Ryu:2006bv, Ryu:2006ef}.
For the curve $\mathcal{C}$ to have the minimal area, $u=u(x)$ and $x=x(u)$ along $\mathcal{C}$ should be single-valued.
Thus, starting from $(u,x)=(\infty,-\frac{L}{2})$, the curve $\mathcal{C}$ extends monotonically away from the subsystem $\mathcal{S}$ until it reaches its deepest point, denoted by $(u_*,x_*)$.
From this point, $\mathcal{C}$ approaches $\mathcal{S}$ monotonically and terminates at $(\infty,\frac{L}{2})$.
In fact, $x_*=0$, since the metric is invariant under the reflection $x\rightarrow -x$.
If one cannot notice any change in the metric after the reflection, then one cannot tell the difference in the shape of the extremal curve after the reflection either.
Therefore, the curve $\mathcal{C}$ is symmetric about the $u$-axis at $x=0$ and has the deepest point at $(u_*,0)$.
With these conditions, the area $A$ of $\mathcal{C}$ is given by
\begin{align}
\label{D3area_general}
	A
	=
		\int \sqrt{g_{uu} du^2 + g_{xx} dx^2}
	=
		2 r_0^2
		\int_{u_*}^\infty du\,
		u^3
		\sqrt{ \frac{1}{u^4 N(u)} + \left( \frac{dx}{du} \right)^2 }.
\end{align}
The variational principle leads to
\begin{align}
	u^3 \left[
			\frac{1}{u^4 N(u)} + \left( \frac{dx}{du} \right)^2
		\right]^{ - \frac{1}{2} }
	\frac{dx}{du}
	=
		\textrm{constant}
	=
		u_*^3.
\end{align}
The constant is equal to $u_*^3$ because $\frac{du}{dx}=0$ at $(u,x)=(u_*,0)$.
Substituting the solution for $\left(\frac{dx}{du}\right)^2$ into \eqref{D3area_general}, we find that the minimal area of $\mathcal{C}$ is given by
\begin{align}
\label{D3area_A_and_u_star}
	A_\textrm{min}
	&=
		2 r_0^2\,
		\lim_{\eta\rightarrow 0}
		\int_{u_*}^\frac{1}{\eta} du\,
			\frac{u}{\sqrt{N(u)}}
			\left(
				1 - \frac{u_*^6}{u^6}		
			\right)^{-\frac{1}{2}}.
\end{align}
The depth $u_*$ is determined by $L$, which can be written as
\begin{align}
\label{D3area_L_and_u_star}
	L
	=
		\int^{\frac{L}{2}}_{-\frac{L}{2}} dx
	=
		2
		\int^\infty_{u_*} du\,
		\frac{1}{u^2}
		\left[ 
			\left( \frac{u^6}{u_*^6}-1 \right)
			N(u)
		\right]^{-\frac{1}{2}}
\end{align}
We are interested in the high-energy limit, where $u \gg 1$.
Hence we expect $u_* \gg 1$ and expand $A_\textrm{min}$ and $L$ in powers of $s = u_*^{-1}$.
However, $u_*$ is not a freely adjustable parameter because it is determined by the variational principle after the background geometry and the shape of the subsystem $S$ are given.
Therefore, we need another assumption on $\mathcal{S}$.
First we restrict ourselves to $s \ll 1$ and expand the integrand in \eqref{D3area_L_and_u_star} in powers of $s$.
After integration, $L$ is given by
\begin{align}
\label{D3_L_in_s}
\begin{split}
	L
	&=
		\frac{ 4\pi^{\frac{3}{2}} }
			{ \sqrt{3}\,
				\Gamma\left(\frac{1}{6}\right)
				\Gamma\left(\frac{1}{3}\right) }\,
		s
	+
		\left[
			\frac{ \Gamma\left(\frac{1}{6}\right)
					\Gamma\left(\frac{1}{3}\right) }
				{ 30\sqrt{\pi} }
			+
			\frac{ 8 \Gamma\left(\frac{1}{6}\right)
					\Gamma\left(\frac{1}{3}\right) }
				{ 45\sqrt{\pi} }\,
			\delta^2
			+
			\frac{ 281\, \Gamma\left(\frac{1}{6}\right)
						\Gamma\left(\frac{1}{3}\right) }
				{ 75600 \sqrt{\pi} }\,
			\delta^2\epsilon^2
		\right]
		s^5
	\\ &\hphantom{=}
	-
		\left[
			\frac{ 128\pi^{\frac{3}{2}} }
				{ 21\sqrt{3}\,
					\Gamma\left(\frac{1}{6}\right)
					\Gamma\left(\frac{1}{3}\right)
				}\, \delta^2
			+
			\frac{ 187\pi^{\frac{3}{2}} }
				{ 735\sqrt{3}\,
					\Gamma\left(\frac{1}{6}\right)
					\Gamma\left(\frac{1}{3}\right)
				}\, \delta^2\epsilon^2
		\right]
		s^7
	+
		\left[
			\frac{1}{6}
			+ \frac{16}{9}\,\delta^2
			- \frac{31}{840}\, \delta^2\epsilon^2
		\right]
		s^9
	+
		\mathcal{O} (s^{11}).
\end{split}
\end{align}
Since there is no constant or negative-power terms in $s$, we conclude that the subsystem $\mathcal{S}$ is assumed to be small, so $L \ll 1$.
In the same way, $A_\textrm{min}$ is expressed in powers of $s$ as
\begin{align}
\label{D3minA_in_s}
\nonumber
	\frac{A_\textrm{min}}{ 2 r_0^2 } &=
		\lim_{\eta\rightarrow 0}
			\left[
				\frac{1}{2 \eta^2}
				-
				\frac{ \pi^\frac{3}{2} }
					{ \sqrt{3}\, \Gamma\left(\frac{1}{6}\right)
								\Gamma\left(\frac{1}{3}\right) }
			\right]
			\frac{1}{s^2}
		+
		\left[
			\frac{ \Gamma\left(\frac{1}{6}\right)
					\Gamma\left(\frac{1}{3}\right) }
				{ 24\pi^\frac{1}{2} }
			+
			\frac{ 2 \Gamma\left(\frac{1}{6}\right)
					\Gamma\left(\frac{1}{3}\right) }
				{ 9\pi^\frac{1}{2} }\,
			\delta^2
			+
			\frac{ 281\, \Gamma\left(\frac{1}{6}\right)
						\Gamma\left(\frac{1}{3}\right) }
				{ 60480\pi^\frac{1}{2} }
			\delta^2\epsilon^2
		\right]
		s^2
	\\
\begin{split}
	&\hphantom{=} \quad
		-
		\left[
			\frac{ 16 \pi^\frac{3}{2} }
				{3\sqrt{3}\, \Gamma\left(\frac{1}{6}\right)
							\Gamma\left(\frac{1}{3}\right) }\,
			\delta^2
			+
			\frac{ 187 \pi^\frac{3}{2} }
				{ 840\sqrt{3}\, \Gamma\left(\frac{1}{6}\right)
							\Gamma\left(\frac{1}{3}\right)}\,
			\delta^2\epsilon^2
		\right]
		s^4
	\\ &\hphantom{=} \quad
		+
		\left[
			\frac{1}{8}
			+
			\frac{4}{3}\, \delta^2
			-
			\frac{31}{1120}\, \delta^2\epsilon^2
		\right]
		s^6
		+
		\mathcal{O} ( s^8 ).
\end{split}
\end{align}
We denote the $s^{-2}$-order term by
\begin{align}
	A_\textrm{ground} =
		2 r_0^2\, \lim_{\eta\rightarrow 0}
			\left[
				\frac{1}{2 \eta^2}
				-
				\frac{ \pi^\frac{3}{2} }
					{ \sqrt{3}\, \Gamma\left(\frac{1}{6}\right)
								\Gamma\left(\frac{1}{3}\right) }
			\right]
			\frac{1}{s^2}.
\end{align}
Actually, this term is reproduced if we repeat the calculation with $N(u)=1$.
In other words, it is the minimal area of $\mathcal{C}$ provided that the background is purely a hyperscaling-violating geometry ($\alpha=2$), where the black brane mass, chemical potential, and $SU(2)$ gauge fields all vanish.
Since we are interested in how these factors change the entanglement entropy $S$ of the subsystem $\mathcal{S}$, we take $A_\textrm{ground}$ as the ground state entanglement entropy.
We define
\begin{align}
	\Delta S
	\equiv
		S - S_\textrm{ground}
	=
		\frac{2\pi}{\kappa_D^2}
		\left(
			A_\textrm{min}
			-
			A_\textrm{ground}
		\right).
\end{align}
As a final step, we substitute $s$ in $\Delta S$ with $L$. 
According to the holographic dictionary, the holographic entanglement entropy $\Delta S$ corresponds to the entanglement entropy of the subsystem $\mathcal{S}$.
Therefore, it should be expressed in terms of parameters defined on the $u=\infty$ boundary, which is the system size $L$ in this case.
Using \eqref{D3_L_in_s}, we write $s$ in terms $L$ as
\begin{align}
\nonumber
	s &=
		\frac{ \sqrt{3}\,
				\Gamma\left(\frac{1}{6}\right)
				\Gamma\left(\frac{1}{3}\right) }
			{ 4\pi^{\frac{3}{2}} }\,
		L
		-
		\left[
			\frac{ 9 \Gamma\left(\frac{1}{6}\right)^7
					\Gamma\left(\frac{1}{3}\right)^7 }
				{ 40960\pi^\frac{19}{2} }
			+
			\frac{ 3 \Gamma\left(\frac{1}{6}\right)^7
					\Gamma\left(\frac{1}{3}\right)^7 }
				{ 2560\pi^\frac{19}{2} }\,
			\delta^2
			+
			\frac{ 281\,
					\Gamma\left(\frac{1}{6}\right)^7
					\Gamma\left(\frac{1}{3}\right)^7}
				{ 11468800\pi^\frac{19}{2} }\,
			\delta^2\epsilon^2
		\right]
		L^5
	\\ &\hphantom{=}
		+
		\left[
			\frac{ 9\sqrt{3}\,
					\Gamma\left(\frac{1}{6}\right)^7
					\Gamma\left(\frac{1}{3}\right)^7 }
				{ 3584\pi^\frac{21}{2} }\, \delta^2
			+
			\frac{ 1683\sqrt{3}\,
					\Gamma\left(\frac{1}{6}\right)^7
					\Gamma\left(\frac{1}{3}\right)^7 }
				{ 16056320\pi^\frac{21}{2} }\,\delta^2\epsilon^2
		\right]
		L^7
		+
		\mathcal{O}(L^{9}).
\end{align}
Finally we have $\Delta S$ in powers of $L$, which is given by
\begin{align}
\begin{split}
	\frac{\kappa_3^2}{4\pi r_0^2}\,
	\Delta S
	&=
		\left[
			\frac{ \Gamma\left(\frac{1}{6}\right)^3
					\Gamma\left(\frac{1}{3}\right)^3}
				{ 128 \pi^\frac{7}{2} }
			+
			\frac{ \Gamma\left(\frac{1}{6}\right)^3
					\Gamma\left(\frac{1}{3}\right)^3}
				{ 24 \pi^\frac{7}{2} }\,
			\delta^2
			+
			\frac{ 281\, \Gamma\left(\frac{1}{6}\right)^3
							\Gamma\left(\frac{1}{3}\right)^3}
				{ 322560 \pi^\frac{7}{2} }\,
			\delta^2\epsilon^2
		\right]
		L^2
	\\ &\hphantom{=}
		-
		\left[
			\frac{ \sqrt{3}\, \Gamma\left(\frac{1}{6}\right)^3
								\Gamma\left(\frac{1}{3}\right)^3}
				{ 16 \pi^\frac{9}{2} }\,
			\delta^2
			+
			\frac{ 187 \sqrt{3}\, \Gamma\left(\frac{1}{6}\right)^3
									\Gamma\left(\frac{1}{3}\right)^3}
				{ 71680 \pi^\frac{9}{2} }\,
			\delta^2\epsilon^2
		\right]
		L^4
	\\ &\hphantom{=}
		+
		\left[
			\frac{ 27\, \Gamma\left(\frac{1}{6}\right)^6
						\Gamma\left(\frac{1}{3}\right)^6}
				{ 32768 \pi^9 }
			-
			\frac{ 3 \sqrt{3}\, \Gamma\left(\frac{1}{6}\right)^9
								\Gamma\left(\frac{1}{3}\right)^9}
				{ 655360 \pi^\frac{23}{2} }
		\right.
		\\ &\hphantom{=} \qquad
		\left.
			+
			\left\{
				\frac{ 9\, \Gamma\left(\frac{1}{6}\right)^6
							\Gamma\left(\frac{1}{3}\right)^6}
				{ 1024 \pi^9 }
				-
				\frac{ \sqrt{3}\, \Gamma\left(\frac{1}{6}\right)^9
									\Gamma\left(\frac{1}{3}\right)^9}
				{ 20480 \pi^\frac{23}{2} }
			\right\}\,
			\delta^2
		\right.
		\\ &\hphantom{=} \qquad
		\left.
			-
			\left\{
				\frac{ 837\,
							\Gamma\left(\frac{1}{6}\right)^6
							\Gamma\left(\frac{1}{3}\right)^6}
				{ 4587520 \pi^9 }
				+
				\frac{ 281\,
							\Gamma\left(\frac{1}{6}\right)^9
							\Gamma\left(\frac{1}{3}\right)^9}
				{ 91750400\sqrt{3} \pi^\frac{23}{2} }
			\right\}\,
			\delta^2\epsilon^2
		\right]
		L^6
		+
		\mathcal{O} (L^8).
\end{split}
\end{align}
\subsubsection{The first law of entanglement entropy}
It is suggested that changes in energy and entanglement entropy of a small system have a relation analogous to the first law of thermodynamics.
The entanglement temperature $\mathcal{T}$ is defined by
\begin{align}
	\mathcal{T}
	=
		\lim_{L \rightarrow 0} \frac{ \Delta E }{ \Delta S }.
\end{align}
Moreover, $\mathcal{T}$ is expected to have a universality: it depends only on the shape of the system and the dimensionality of the boundary spacetime it lies on.

We may think of the energy of this system, which is given by
\begin{align}
	\Delta E
	=
		E - E_\textrm{ground}
	=
		\int_\textrm{line} dx\,
			\langle T_{tt} \rangle,
\end{align}
where $\langle T_{tt} \rangle$ is the energy density on the boundary.
We assume that $\langle T_{tt} \rangle$ is constant and it is related to the black brane mass $M$ by
\begin{align}
	\langle T_{tt} \rangle =
		\frac{(D-2)M}{2\kappa_D^2}
	=
		\frac{M}{2\kappa_3^2},
\end{align}
which leads to
\begin{align}
	\Delta E = \frac{M}{2\kappa_3^2}\, L.
\end{align}
The black brane mass is obtained from the coefficient of the $u^{-4}$ term in the function $N(u)$ by locating the horizon at $u=1$, as shown in \eqref{metric_components_without_backreaction} - \eqref{black_brane_mass_from_N_2}.
In isotropic phase, where $\epsilon = 0$, $M$ is given by
\begin{align}
	M_\textrm{iso} = r_0^4 \left( 1 + \frac{16}{3}\delta^2 \right).
\end{align}
In anisotropic phase, it becomes
\begin{align}
	M_\textrm{aniso}
	=
		r_0^4
			\left(
				1
				+ \frac{16}{3}\, \delta^2
				+ \frac{281}{2520}\, \delta^2\epsilon^2
			\right).
\end{align}

There are three distinct contributions to $\Delta E$ and $\Delta S$.
The black brane provides the first contribution, which yields terms independent of $\delta$ and $\epsilon$.
These terms are denoted by $\Delta_0$.
The second contribution arises from the chemical potential, incoded in the temporal part of $SU(2)$ gauge field.
It gives rise to $\delta^2$-order terms, which are denoted by $\Delta_\delta$.
The third one occurs when the vector order $\epsilon$ appears and breaks a symmetry.
The resulting terms in order of $\delta^2\epsilon^2$ are denoted by $\Delta_\epsilon$.
To make these notations explicit,
\begin{align}
\begin{split}
	\Delta E
	&=
		\Delta_0 E
		+
		\delta^2 \Delta_\delta E
		+
		\delta^2\epsilon^2 \Delta_\epsilon E
	\\	
	&=
		\frac{r_0^4}{2\kappa_3^2}\, L
		+
		\delta^2 \left( \frac{ 8 r_0^4 }{ 3 \kappa_3^2 }\, L \right)
		+
		\delta^2\epsilon^2
			\left(
				\frac{ 281\, r_0^4 }{ 5040 \kappa_3^2 }\, L
			\right)
\end{split}
\end{align}
and
\begin{align}
\begin{split}
	\Delta S
	&=
		\Delta_0 S
		+
		\delta^2 \Delta_\delta S
		+
		\delta^2\epsilon^2 \Delta_\epsilon S
	\\
	&=
		\frac{ r_0^2\, \Gamma\left(\frac{1}{6}\right)^3
						\Gamma\left(\frac{1}{3}\right)^3 }
			{ 32 \pi^\frac{5}{2} \kappa_3^2 }\,
		L^2
		+
		\delta^2
			\left[
				\frac{ r_0^2\, \Gamma\left(\frac{1}{6}\right)^3
								\Gamma\left(\frac{1}{3}\right)^3 }
					{ 6 \pi^\frac{5}{2} \kappa_3^2 }\,
				L^2
			\right]
		+
		\delta^2\epsilon^2
			\left[
				\frac{ 281\, r_0^2\, \Gamma\left(\frac{1}{6}\right)^3
										\Gamma\left(\frac{1}{3}\right)^3 }
						{ 80640 \pi^\frac{5}{2} \kappa_3^2 }\,
				L^2
			\right]
		+
		\mathcal{O} (L^4).
\end{split}
\end{align}
If the analogy with the first law of thermodynamics is consistent, we should have 
\begin{align}
	\mathcal{T}
	=
		\lim_{L\rightarrow 0}
			\frac{\Delta E}{\Delta S}
	=
		\lim_{L\rightarrow 0}
			\frac{\Delta_0 E}{\Delta_0 S}
	=
		\lim_{L\rightarrow 0}
			\frac{\Delta_\delta E}{\Delta_\delta S}
	=
		\lim_{L\rightarrow 0}
			\frac{\Delta_\epsilon E}{\Delta_\epsilon S}.
\end{align}
Indeed, we find
\begin{align}
	\mathcal{T}
	=
	\frac{ 16 \pi^\frac{5}{2} r_0^2 }
		{ \Gamma\left(\frac{1}{6}\right)^3
			\Gamma\left(\frac{1}{3}\right)^3}\,
	\frac{1}{L}
	\approx
	0.084412\, \frac{r_0^2}{L}.
\end{align}
It also shows a universality even in the anisotropic phase.
\subsubsection{Scaling behavior near critical point}
In Sec.\ref{section_introduction}, we discussed that when the black brane temperature is below the critical temperature $T_c$, a vector order $\epsilon$ appears and the system undergoes a phase transition from isotropic to anisotropic phase.
We see that the change in the entanglement entropy $\Delta S$ shows a scaling behavior near the critical point.
The black brane temperature is given by
\begin{align}
	T
	=
		\frac{1}{\pi r_0^5}
		\left(
			1
			-
			\frac{8}{3}\, \delta^2
			+
			\frac{23}{1680}\, \delta^2\epsilon^2
		\right),
\end{align}
and the critical temperature is given by
\begin{align}
	T_c
	=
		\frac{1}{\pi r_0^5}
		\left(
			1
			-
			\frac{8}{3}\, \delta^2
		\right).
\end{align}
We can express $\delta^2\epsilon^2$ in terms of $(T-T_c)$.
Then, up to the leading order in $L$, we have
\begin{align}
	\Delta_\epsilon S
	=
		\frac{
			281\,
			\Gamma \left( \frac{1}{6} \right)^3
			\Gamma \left( \frac{1}{3} \right)^3 }
			{
			1104\, \pi^\frac{3}{2} \kappa_3^2 }
		r_0^7\,
		L^2
		\left( T - T_c \right)^\beta
	\approx
		\frac{ 152 }{ \kappa_3^2 }
		\, r_0^7\, L^2
		\left( T - T_c \right)^\beta
		,
\end{align}
with the critical exponent $\beta$ equal to one.
\subsection{$D=4$ case: Strap-shaped subsystems on $xy$-plane}
\subsubsection{Holographic entanglement entropy of a strap lying along $x$-axis}
\begin{figure}[h]
	\centering
	\includegraphics[scale=0.25]{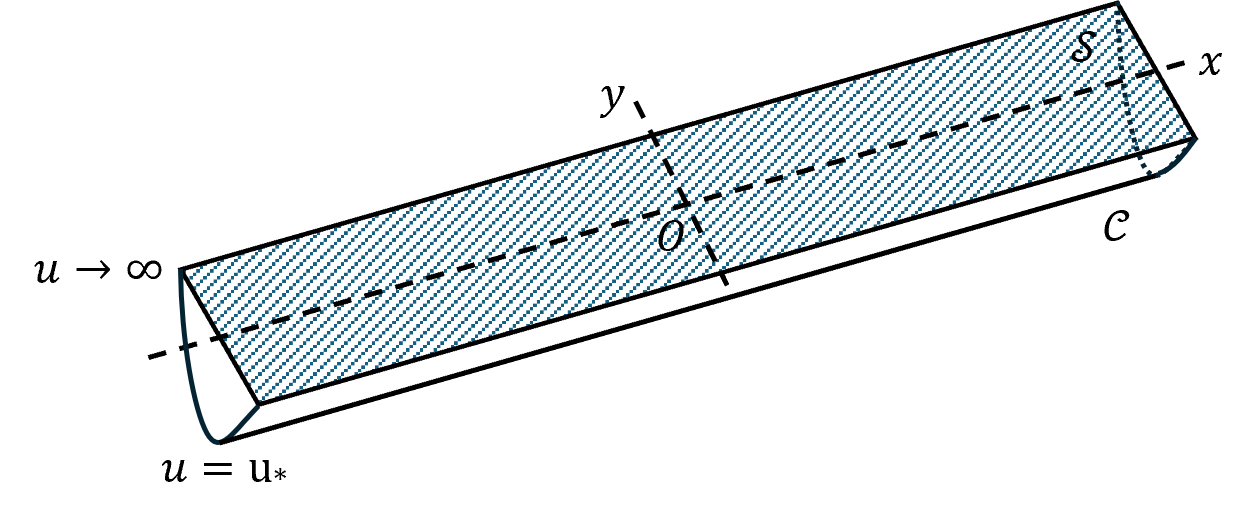}
	\caption{}
	\label{figure_D4_strap_x}
\end{figure}
\noindent
Consider a strap-shaped subsystem $\mathcal{S}$ on the boundary, lying along the $x$-axis at $u\rightarrow\infty$.
It is infinitely long and very narrow so that $-\frac{L}{2} \leq x \leq \frac{L}{2}$ and $-\frac{W}{2} \leq y \leq \frac{W}{2}$, with $L\rightarrow\infty$ and $W \ll 1$.
To obtain the holographic entanglement entropy of $\mathcal{S}$, we consider a curved strap $\mathcal{C}$ that hangs down in the bulk direction from the long sides of $\mathcal{S}$.
Since the strap is infinitely long, the surfaces at $x=\pm\frac{L}{2}$ can be neglected.
Generalizing the discussion given in the previous section, we assume that $\mathcal{C}$ is symmetric and $u=u(x,y)$ is single-valued on $\mathcal{C}$, as shown in Figure \ref{figure_D4_strap_x}.
The deepest line on $C$ is defined by $(u,y)=(u_*,0)$.
With these conditions, the area $A$ of $\mathcal{C}$ is given by
\begin{align}
\begin{split}
	A
	=
		\int \sqrt{g_{xx}}\, dx\, \sqrt{g_{uu} du^2 + g_{yy} dy^2}
	=
		2 r_0 L \int_{u_*}^\infty du\,
			u^3 H J	\sqrt{ \frac{1}{u^4 J^2 N}
	+
		\left( \frac{dy}{du} \right)^2 },
\end{split}
\end{align}
The variational principle leads to
\begin{align}
	u^3 H J
		\left[
			\frac{1}{u^4 J^2 N} + \left( \frac{dy}{du} \right)^2
		\right]^{ - \frac{1}{2} }
	\frac{dy}{du}
	=
		\textrm{constant}
	=
		u_*^3 H_* J_*,
\end{align}
where $H_*=H(u_*)$ and $J_*=J(u_*)$.
The constant is equal to $u_*^3 H_* J_*$ because $\frac{du}{dy}=0$ at $(u,y)=(u_*,0)$.
Substituting the solution for $\left( \frac{dy}{du} \right)^2$ into the integrand, the minimal area of $\mathcal{C}$ is given by
\begin{align}
	A_\textrm{min}
	&=
		2 r_0 L\,
		\lim_{\eta\rightarrow 0}
		\int_{u_*}^\frac{1}{\eta} du\,
			\frac{u H}{\sqrt{N}}
			\left(
				1 - \frac{u_*^6 H_*^2 J_*^2}{u^6 H^2 J^2}		
			\right)^{-\frac{1}{2}}.
\end{align}
The depth $u_*$ is determined by $W$, which can be written as
\begin{align}
\begin{split}
	W
	=
		\int^{\frac{W}{2}}_{-\frac{W}{2}} dy
	=
		2
		\int^\infty_{u_*} du\,
		\frac{1}{u^2 J}
			\left[ 
				\left( \frac{u^6H^2J^2}{u_*^6H_*^2J_*^2}-1 \right) N
			\right]^{-\frac{1}{2}}.
\end{split}
\end{align}
As in the previous section, a small system size leads to a large $u_*$, allowing us to probe the high-energy limit.
In this case, we assumed that the width $W$ of $\mathcal{S}$ is small.
Therefore, $W$ can be expanded in powers of $s=u_*^{-1}$ as
\begin{align}
\begin{split}
	W
	&=
		\frac{ 4\pi^{\frac{3}{2}} }
			{ \sqrt{3}\,
				\Gamma\left(\frac{1}{6}\right)
				\Gamma\left(\frac{1}{3}\right)
			}\,
		s
	+
		\left[
			\frac{ \Gamma\left(\frac{1}{6}\right)
					\Gamma\left(\frac{1}{3}\right) }
				{ 30\sqrt{\pi} }
			+
			\frac{ 8 \Gamma\left(\frac{1}{6}\right)
					\Gamma\left(\frac{1}{3}\right) }
				{ 45\sqrt{\pi} }\,
			\delta^2
			-
			\frac{ 281\, \Gamma\left(\frac{1}{6}\right)
						\Gamma\left(\frac{1}{3}\right) }
				{ 113400 \sqrt{\pi} }\,
			\delta^2\epsilon^2
		\right]
		s^5
	\\ &\hphantom{=}
	-
		\left[
			\frac{ 128\pi^{\frac{3}{2}} }
				{ 21\sqrt{3}\,
					\Gamma\left(\frac{1}{6}\right)
					\Gamma\left(\frac{1}{3}\right)
				}\, \delta^2
			+
			\frac{ 467\pi^{\frac{3}{2}} }
				{ 735\sqrt{3}\,
					\Gamma\left(\frac{1}{6}\right)
					\Gamma\left(\frac{1}{3}\right)
				}\, \delta^2\epsilon^2
		\right]
		s^7
	+
		\left[
			\frac{1}{6}
			+ \frac{16}{9}\,\delta^2
			+ \frac{2659}{11340}\, \delta^2\epsilon^2
		\right]
		s^9
	+
		\mathcal{O} (s^{11}).
\end{split}
\end{align}
From this result, $s$ is written in terms of $W$ as
\begin{align}
\nonumber
	s
	&=
		\frac{ \sqrt{3}\,
				\Gamma\left(\frac{1}{6}\right)
				\Gamma\left(\frac{1}{3}\right) }
			{ 4\pi^{\frac{3}{2}} }\,
		W
	-
		\left[
			\frac{ 9 \Gamma\left(\frac{1}{6}\right)^7
					\Gamma\left(\frac{1}{3}\right)^7 }
				{ 40960\pi^\frac{19}{2} }
			+
			\frac{ 3 \Gamma\left(\frac{1}{6}\right)^7
					\Gamma\left(\frac{1}{3}\right)^7 }
				{ 2560\pi^\frac{19}{2} }\,
			\delta^2
			-
			\frac{ 281\,
					\Gamma\left(\frac{1}{6}\right)^7
					\Gamma\left(\frac{1}{3}\right)^7}
				{ 17203200\pi^\frac{19}{2} }\,
			\delta^2\epsilon^2
		\right]
		W^5
	\\ &\hphantom{=}
	+
		\left[
			\frac{ 9\sqrt{3}\,
					\Gamma\left(\frac{1}{6}\right)^7
					\Gamma\left(\frac{1}{3}\right)^7 }
				{ 3584\pi^\frac{21}{2} }\, \delta^2
			+
			\frac{ 4203\sqrt{3}\,
					\Gamma\left(\frac{1}{6}\right)^7
					\Gamma\left(\frac{1}{3}\right)^7 }
				{ 16056320\pi^\frac{21}{2} }\,\delta^2\epsilon^2
		\right]
		W^7
	+
		\mathcal{O}(W^{9}).
\end{align}
The expansion of $A_\textrm{min}$ is given by
\begin{align}
\begin{split}
	\frac{A_\textrm{min}}{ 2 r_0 L }
	&=
		\frac{A_\textrm{ground}}{2r_0 L}
	+
		\left[
			\frac{ \Gamma\left(\frac{1}{6}\right)
					\Gamma\left(\frac{1}{3}\right) }
				{ 24\pi^\frac{1}{2} }
			+
			\frac{ 2 \Gamma\left(\frac{1}{6}\right)
					\Gamma\left(\frac{1}{3}\right) }
				{ 9\pi^\frac{1}{2} }\,
			\delta^2
			+
			\frac{ 281\, \Gamma\left(\frac{1}{6}\right)
						\Gamma\left(\frac{1}{3}\right) }
				{ 60480\pi^\frac{1}{2} }\,
			\delta^2\epsilon^2
		\right]
		s^2
	\\
	 &\hphantom{=}\quad
	 -
		\left[
			\frac{ 16 \pi^\frac{3}{2} }
				{3\sqrt{3}\, \Gamma\left(\frac{1}{6}\right)
							\Gamma\left(\frac{1}{3}\right) }\,
			\delta^2
			+
			\frac{ 467 \pi^\frac{3}{2} }
				{ 840\sqrt{3}\, \Gamma\left(\frac{1}{6}\right)
							\Gamma\left(\frac{1}{3}\right)}\,
			\delta^2\epsilon^2
		\right]
		s^4
	+
		\left[
			\frac{1}{8}
			+
			\frac{4}{3}\, \delta^2
			+
			\frac{249}{1120}\, \delta^2\epsilon^2
		\right]
		s^6
	+
		\mathcal{O} ( s^8 ),
\end{split}
\end{align}
where
\begin{align*}
	A_\textrm{ground}
	=
		2 r_0 L\,
		\lim_{\eta\rightarrow 0}
			\left[
				\frac{1}{2 \eta^2}
				-
				\frac{ \pi^\frac{3}{2} }
					{ \sqrt{3}\, \Gamma\left(\frac{1}{6}\right)
								\Gamma\left(\frac{1}{3}\right) }
			\right]
			\frac{1}{s^2}
\end{align*}
is the minimal area of $\mathcal{C}$ provided that the background is purely a hyperscaling-violating geometry with $\alpha = \frac{1}{2}$, which is given by
Note that this is the same result as in the previous $D=3$, $\alpha=2$ case.
Finally we obtain
\begin{align}
\label{StrapAlongX_minArea}
	\frac{\kappa_4^2}{4\pi r_0L}\, \Delta S
	&=
		\frac{ A_\textrm{min} - A_\textrm{ground} }{ 2 r_0 L }
	\\
\begin{split}
	&=
		\left[
			\frac{ \Gamma\left(\frac{1}{6}\right)^3
					\Gamma\left(\frac{1}{3}\right)^3}
				{ 128 \pi^\frac{7}{2} }
			+
			\frac{ \Gamma\left(\frac{1}{6}\right)^3
					\Gamma\left(\frac{1}{3}\right)^3}
				{ 24 \pi^\frac{7}{2} }\,
			\delta^2
			+
			\frac{ 281\, \Gamma\left(\frac{1}{6}\right)^3
							\Gamma\left(\frac{1}{3}\right)^3}
				{ 322560 \pi^\frac{7}{2} }\,
			\delta^2\epsilon^2
		\right]
		W^2
	\\ &\hphantom{=}
	-
		\left[
			\frac{ \sqrt{3}\, \Gamma\left(\frac{1}{6}\right)^3
								\Gamma\left(\frac{1}{3}\right)^3}
				{ 16 \pi^\frac{9}{2} }\,
			\delta^2
			+
			\frac{ 467 \sqrt{3}\, \Gamma\left(\frac{1}{6}\right)^3
									\Gamma\left(\frac{1}{3}\right)^3}
				{ 71680 \pi^\frac{9}{2} }\,
			\delta^2\epsilon^2
		\right]
		W^4
	\\ &\hphantom{=}
	+
		\left[
			\frac{ 27\, \Gamma\left(\frac{1}{6}\right)^6
						\Gamma\left(\frac{1}{3}\right)^6}
				{ 32768 \pi^9 }
			-
			\frac{ 3 \sqrt{3}\, \Gamma\left(\frac{1}{6}\right)^9
								\Gamma\left(\frac{1}{3}\right)^9}
				{ 655360 \pi^\frac{23}{2} }
		\right.
	\\ &\hphantom{=} \qquad
		\left.
	+
			\left\{
				\frac{ 9\, \Gamma\left(\frac{1}{6}\right)^6
							\Gamma\left(\frac{1}{3}\right)^6}
				{ 1024 \pi^9 }
				-
				\frac{ \sqrt{3}\, \Gamma\left(\frac{1}{6}\right)^9
									\Gamma\left(\frac{1}{3}\right)^9}
				{ 20480 \pi^\frac{23}{2} }
			\right\}\,
			\delta^2
		\right.
	\\ &\hphantom{=} \qquad
		\left.
	+
			\left\{
				\frac{ 6723\,
							\Gamma\left(\frac{1}{6}\right)^6
							\Gamma\left(\frac{1}{3}\right)^6}
				{ 4587520 \pi^9 }
				-
				\frac{ 281\,
							\Gamma\left(\frac{1}{6}\right)^9
							\Gamma\left(\frac{1}{3}\right)^9}
				{ 550502400\sqrt{3} \pi^\frac{23}{2} }
			\right\}\,
			\delta^2\epsilon^2
		\right]
		W^6
	+
		\mathcal{O} (W^8).
\end{split}
\end{align}
\subsubsection{Holographic entanglement entropy of a strap lying along $y$-axis}
Consider a strap-shaped subsystem $\mathcal{S}$ that lies at $u=\infty$ along the $y$-axis, $-\frac{L}{2} \leq y \leq \frac{L}{2}$ and $-\frac{W}{2} \leq x \leq \frac{W}{2}$.
A curved strap $\mathcal{C}$ hangs down in the bulk direction, from the long sides of $\mathcal{S}$.
The other conditions are similar to those in the previous case.
The surface area of $\mathcal{C}$ is given by
\begin{align}
\begin{split}
	A
	=
		2 \int \sqrt{g_{yy}}\, dy\, \sqrt{g_{uu} du^2 + g_{xx} dx^2}
	=
		2 r_0 L \int_{u_*}^\infty du\,
			u^3 H J
			\sqrt{
				\frac{1}{u^4 H^2 N}
				+
				\left( \frac{dx}{du} \right)^2
			},
\end{split}
\end{align}
where $u_*$ is the minimum value of $u$ that $\mathcal{C}$ may have.
The variational principle leads to
\begin{align}
	u^3 H J
		\left[
			\frac{1}{u^4 H^2 N} + \left( \frac{dx}{du} \right)^2
		\right]^{ - \frac{1}{2} }
	\frac{dx}{du}
	=
		u_*^3 H_* J_*,
\end{align}
where $H_*=H(u_*)$ and $J_*=J(u_*)$.
The minimal area of $\mathcal{C}$ is given by
\begin{align}
	A_\textrm{min}
	&=
		2 r_0 L
		\lim_{\eta\rightarrow\infty}
		\int^\frac{1}{\eta}_{u_*} du\,
			\frac{u J}{\sqrt{N}}
			\left(
				1 - \frac{u_*^6 H_*^2 J_*^2}{u^6 H^2 J^2}		
			\right)^{-\frac{1}{2}}.
\end{align}
The system size $W$ can be written as
\begin{align}
\begin{split}
	W
	=
		\int^{\frac{W}{2}}_{-\frac{W}{2}} dx
	=
		2\lim_{\delta\rightarrow 0}
			\int^{\frac{1}{\delta}}_{u_*} du\,
			\frac{1}{u^2 H}
				\left[ 
					\left( \frac{u^6H^2J^2}{u_*^6H_*^2J_*^2}-1 \right) N
				\right]^{-\frac{1}{2}}.
\end{split}
\end{align}
In powers of $s = u_*^{-1} \ll 1$, it is expanded as
\begin{align}
\begin{split}
	W
	&=
		\frac{ 4\pi^{\frac{3}{2}} }
			{ \sqrt{3}\,
				\Gamma\left(\frac{1}{6}\right)
				\Gamma\left(\frac{1}{3}\right) }\,
		s
		+
		\left[
			\frac{ \Gamma\left(\frac{1}{6}\right)
					\Gamma\left(\frac{1}{3}\right) }
				{ 30\sqrt{\pi} }
			+
			\frac{ 8 \Gamma\left(\frac{1}{6}\right)
					\Gamma\left(\frac{1}{3}\right) }
				{ 45\sqrt{\pi} }\,
			\delta^2
			+
			\frac{ 281\, \Gamma\left(\frac{1}{6}\right)
						\Gamma\left(\frac{1}{3}\right) }
				{ 75600 \sqrt{\pi} }\,
			\delta^2\epsilon^2
		\right]
		s^5
	\\ &\hphantom{=}
	-
		\left[
			\frac{ 128\pi^{\frac{3}{2}} }
				{ 21\sqrt{3}\,
					\Gamma\left(\frac{1}{6}\right)
					\Gamma\left(\frac{1}{3}\right)
				}\, \delta^2
			+
			\frac{ 187\pi^{\frac{3}{2}} }
				{ 735\sqrt{3}\,
					\Gamma\left(\frac{1}{6}\right)
					\Gamma\left(\frac{1}{3}\right)
				}\, \delta^2\epsilon^2
		\right]
		s^7
	+
		\left[
			\frac{1}{6}
			+
			\frac{16}{9}\, \delta^2
			-
			\frac{31}{840}\, \delta^2\epsilon^2
		\right]
	+
		\mathcal{O} (s^{11}).
\end{split}
\end{align}
From this result, we get
\begin{align}
\nonumber
	s
	&=
		\frac{ \sqrt{3}\,
				\Gamma\left(\frac{1}{6}\right)
				\Gamma\left(\frac{1}{3}\right) }
			{ 4\pi^{\frac{3}{2}} }\,
		W
	-
		\left[
			\frac{ 9 \Gamma\left(\frac{1}{6}\right)^7
					\Gamma\left(\frac{1}{3}\right)^7 }
				{ 40960\pi^\frac{19}{2} }
			+
			\frac{ 3 \Gamma\left(\frac{1}{6}\right)^7
					\Gamma\left(\frac{1}{3}\right)^7 }
				{ 2560\pi^\frac{19}{2} }\,
			\delta^2
			+
			\frac{ 281\,
					\Gamma\left(\frac{1}{6}\right)^7
					\Gamma\left(\frac{1}{3}\right)^7}
				{ 11468800\pi^\frac{19}{2} }\,
			\delta^2\epsilon^2
		\right]
		W^5
	\\ &\hphantom{=}
	+
		\left[
			\frac{ 9\sqrt{3}\,
					\Gamma\left(\frac{1}{6}\right)^7
					\Gamma\left(\frac{1}{3}\right)^7 }
				{ 3584\pi^\frac{21}{2} }\, \delta^2
			+
			\frac{ 1683\sqrt{3}\,
					\Gamma\left(\frac{1}{6}\right)^7
					\Gamma\left(\frac{1}{3}\right)^7 }
				{ 16056320\pi^\frac{21}{2} }\,\delta^2\epsilon^2
		\right]
		W^7
		+
		\mathcal{O}(W^{9}).
\end{align}
The expansion of $A_\textrm{min}$ is given by
\begin{align}
\begin{split}
	\frac{A}{ 2 r_0 L }
	&=
		\frac{A_\textrm{ground}}{2 r_0 L}
	+
		\left[
			\frac{ \Gamma\left(\frac{1}{6}\right)
					\Gamma\left(\frac{1}{3}\right) }
				{ 24\pi^\frac{1}{2} }
			+
			\frac{ 2 \Gamma\left(\frac{1}{6}\right)
					\Gamma\left(\frac{1}{3}\right) }
				{ 9\pi^\frac{1}{2} }\,
			\delta^2
			+
			\frac{ 281\, \Gamma\left(\frac{1}{6}\right)
						\Gamma\left(\frac{1}{3}\right) }
				{ 60480\pi^\frac{1}{2} }\,
			\delta^2\epsilon^2
		\right]
		s^2
	\\ &\hphantom{=} \quad
	-
		\left[
			\frac{ 16 \pi^\frac{3}{2} }
				{3\sqrt{3}\, \Gamma\left(\frac{1}{6}\right)
							\Gamma\left(\frac{1}{3}\right) }\,
			\delta^2
			+
			\frac{ 187 \pi^\frac{3}{2} }
				{ 840\sqrt{3}\, \Gamma\left(\frac{1}{6}\right)
							\Gamma\left(\frac{1}{3}\right)}\,
			\delta^2\epsilon^2
		\right]
		s^4
	+
		\left[
			\frac{1}{8}
			+
			\frac{4}{3}\, \delta^2
			+
			\frac{249}{1120}\, \delta^2\epsilon^2
		\right]
		s^6
	+
		\mathcal{O} ( s^8 ).
\end{split}
\end{align}
Finally, the entanglement entropy change $\Delta S$ is given by
\begin{align}
\label{StrapAlongY_minArea}
	\frac{\kappa_4^2}{4\pi r_0L} \Delta S
	&=
		\frac{ A_\textrm{min} - A_\textrm{ground} }{ 2 r_0 L }
	\\
\begin{split}
	&=
		\left[
			\frac{ \Gamma\left(\frac{1}{6}\right)^3
					\Gamma\left(\frac{1}{3}\right)^3}
				{ 128 \pi^\frac{7}{2} }
			+
			\frac{ \Gamma\left(\frac{1}{6}\right)^3
					\Gamma\left(\frac{1}{3}\right)^3}
				{ 24 \pi^\frac{7}{2} }\,
			\delta^2
			+
			\frac{ 281\, \Gamma\left(\frac{1}{6}\right)^3
							\Gamma\left(\frac{1}{3}\right)^3}
				{ 322560 \pi^\frac{7}{2} }\,
			\delta^2\epsilon^2
		\right]
		W^2
	\\ &\hphantom{=}
	-
		\left[
			\frac{ \sqrt{3}\, \Gamma\left(\frac{1}{6}\right)^3
								\Gamma\left(\frac{1}{3}\right)^3}
				{ 16 \pi^\frac{9}{2} }\,
			\delta^2
			+
			\frac{ 467 \sqrt{3}\, \Gamma\left(\frac{1}{6}\right)^3
									\Gamma\left(\frac{1}{3}\right)^3}
				{ 71680 \pi^\frac{9}{2} }\,
			\delta^2\epsilon^2
		\right]
		W^4
	\\ &\hphantom{=}
		+
		\left[
			\frac{ 27\, \Gamma\left(\frac{1}{6}\right)^6
						\Gamma\left(\frac{1}{3}\right)^6}
				{ 32768 \pi^9 }
			-
			\frac{ 3 \sqrt{3}\, \Gamma\left(\frac{1}{6}\right)^9
								\Gamma\left(\frac{1}{3}\right)^9}
				{ 655360 \pi^\frac{23}{2} }
		\right.
	\\ &\hphantom{=} \qquad
		\left.
	+
			\left\{
				\frac{ 9\, \Gamma\left(\frac{1}{6}\right)^6
							\Gamma\left(\frac{1}{3}\right)^6}
				{ 1024 \pi^9 }
				-
				\frac{ \sqrt{3}\, \Gamma\left(\frac{1}{6}\right)^9
									\Gamma\left(\frac{1}{3}\right)^9}
				{ 20480 \pi^\frac{23}{2} }
			\right\}\,
			\delta^2
		\right.
	\\ &\hphantom{=} \qquad
		\left.
	+
			\left\{
				\frac{ 6723\,
							\Gamma\left(\frac{1}{6}\right)^6
							\Gamma\left(\frac{1}{3}\right)^6}
				{ 4587520 \pi^9 }
				-
				\frac{ 281\,
							\Gamma\left(\frac{1}{6}\right)^9
							\Gamma\left(\frac{1}{3}\right)^9}
				{ 91750400\sqrt{3} \pi^\frac{23}{2} }
			\right\}\,
			\delta^2\epsilon^2
		\right]
		W^6
	+
		\mathcal{O} (W^8).
\end{split}
\end{align}
\subsubsection{The first law of entanglement entropy}
The energy change of a strap is given by
\begin{align}
	\Delta E =
		\int_\textrm{strap} dxdy\,
			\langle T_{tt} \rangle
	=
		\frac{M}{\kappa_4^2}\, L W,
\end{align}
where $\langle T_{tt} \rangle$ is the energy density on the boundary, assumed to be constant, and $M$ is the black brane mass.
In isotropic and anisotropic phases, $M$ is given by
\begin{align}
	M_\textrm{iso}
	=
		r_0^4 \left( 1 + \frac{16}{3}\delta^2 \right)
	\quad
	\textrm{and}
	\quad
	M_\textrm{aniso}
	=
		r_0^4
			\left(
				1
				+ \frac{16}{3} \delta^2
				+ \frac{281}{2520}\, \delta^2\epsilon^2
			\right).
\end{align}
We have two straps, one lying parallel to the vector order and the other perpendicular to it.
The changes in the entanglement entropies of them start to differ at order $W^4$ and $\delta^2\epsilon^2$.
However, they can be ignored when computing the entanglement temperature, since $W\rightarrow 0$ and the leading order in $\Delta S$ is $W^2$.
We have
\begin{align}
\begin{split}
	\Delta E
	&=
		\Delta_0 E
	+
		\delta^2 \Delta_\delta E
	+
		\delta^2\epsilon^2 \Delta_\epsilon E
	\\
	&=
		\frac{r_0^4 L}{\kappa_4^2}\, W
	+
		\delta^2 \left( \frac{ 16 r_0^4 L }{ 3 \kappa_4^2 }\, W \right)
	+
		\delta^2\epsilon^2
			\left(
				\frac{281 r_0^4 L }{2520\kappa_4^2 }\, W
			\right)
\end{split}
\end{align}
and
\begin{align}
\begin{split}
	\Delta S
	&=
		\Delta_0 S
	+
		\delta^2 \Delta_\delta S
	+
		\delta^2\epsilon^2 \Delta_\epsilon S
	\\
	&=
		\frac{ r_0 L \, \Gamma\left(\frac{1}{6}\right)^3
						\Gamma\left(\frac{1}{3}\right)^3 }
			{ 32 \pi^\frac{5}{2} \kappa_4^2 }\,
		W^2
	+
		\delta^2
			\left[
				\frac{ r_0 L\, \Gamma\left(\frac{1}{6}\right)^3
								\Gamma\left(\frac{1}{3}\right)^3 }
					{ 6 \pi^\frac{5}{2} \kappa_4^2 }\,
				W^2
			\right]
	+
		\delta^2\epsilon^2
			\left[
				\frac{ 281 r_0 L\, \Gamma\left(\frac{1}{6}\right)^3
										\Gamma\left(\frac{1}{3}\right)^3 }
						{ 80640 \pi^\frac{5}{2} \kappa_4^2 }\,
				W^2
			\right]
		+
		\mathcal{O} (W^4).
\end{split}
\end{align}
As expected, the entanglement temperatures $\mathcal{T}$ of both straps are given by
\begin{align}
	\mathcal{T}
	=
		\lim_{W\rightarrow 0} \frac{\Delta E}{\Delta S}
	=
		\lim_{W\rightarrow 0} \frac{\Delta_0 E}{\Delta_0 S}
	=
		\lim_{W\rightarrow 0}
			\frac{\Delta_\delta E}{\Delta_\delta S}
	=
		\lim_{W\rightarrow 0}
			\frac{\Delta_\epsilon E}{\Delta_\epsilon S},
\end{align}
with
\begin{align}
	\mathcal{T}
	=
	\frac{ 32 \pi^\frac{5}{2} r_0^3 }
		{ \Gamma\left(\frac{1}{6}\right)^3
			\Gamma\left(\frac{1}{3}\right)^3}\,
	\frac{1}{W}
	\approx
	0.168823\, \frac{r_0^3}{W}.
\end{align}
Again, we see that $\mathcal{T}$ shows a universality, depending on the inverse of the system size $W$.
Whether the subsystem is lying parallel or perpendicular to the vector order is irrelevant.
\subsubsection{Scaling behavior near critical point}
The black brane temperature and the critical temperature are given by
\begin{align}
	T
	=
		\frac{1}{\pi r_0^2}
		\left(
			1
			-
			\frac{8}{3}\, \delta^2
			+
			\frac{127}{6720}\, \delta^2\epsilon^2
		\right)
	\qquad
	\textrm{and}
	\qquad
	T_c
	=
		\frac{1}{\pi r_0^2}
		\left(
			1
			-
			\frac{8}{3}\, \delta^2
		\right).
\end{align}
Substituting $\delta^2\epsilon^2$ in $\Delta S$ with $(T-T_c)$, up to the leading order in $W$, we have
\begin{align}
	\Delta_\epsilon S
	=
		\frac{
			281\,
			\Gamma \left( \frac{1}{6} \right)^3
			\Gamma \left( \frac{1}{3} \right)^3 }
			{
			1524\, \pi^\frac{3}{2} \kappa_4^2 }
		r_0^3\,
		L
		W^2
		\left( T - T_c \right)^\beta
	\approx
		\frac{ 110 }{ \kappa_4^2 }\,
		r_0^3\,
		L
		W^2
		\left( T - T_c \right)^\beta
		,
\end{align}
with the critical exponent $\beta$ is equal to one.
The same result is obtained for both straps, because the first difference in $\Delta_\epsilon S$ appears at order $W^4$.
\section*{Acknowledgement}
J.H.O thanks his W.J. and Y.J. and he thanks God.
This work was supported by the National Research Foundation of Korea(NRF) grant funded by the Korea government(MSIT). (No.2021R1F1A1047930). 
\section*{Appendices}
\appendix
\section{Solutions for Einstein-Scalar-U(1)$\times$SU(2) Action}
The Einstein-Scalar-U(1)$\times$SU(2) that theory we study is motivated by the previous holographic model, Einstein-Scalar-U(1) theory, in which there are two U(1) gauge fields.
Hence, we introduce briefly the solutions for this model obtained in \cite{Alishahiha:2012qu}.
The action is given by
\begin{align}
    S = - \frac{1}{16\pi G} \int d^{d+2}x \sqrt{-g}
    	\left[
    		R - \frac{1}{2}(\partial\phi)^2 + V_0 e^{\gamma\phi}
    		- \frac{1}{4} \frac{e^{\lambda_F \phi}}{(g_F)^2} F_{\mu\nu}F^{\mu\nu}
    		- \frac{1}{4} \frac{e^{\lambda_G \phi}}{(g_G)^2} G_{\mu\nu}G^{\mu\nu}
    	\right],
\end{align}
where the ansatz for the metric is
\begin{align}
    ds^2 = r^{2\alpha} \left( -r^{2z} f(r) dt^2 + \frac{dr^2}{r^2 f(r)} + r^2 d{\Vec{x}_d}^2 \right)
\end{align}
and the ansatz for the scalar field $\phi$ and the U(1) gauge fields are
\begin{align}
	\phi = \phi (r),
	\qquad
	F_{rt}\neq 0,
	\qquad
	G_{rt}\neq 0,
	\quad
	(F_{\mu\nu}=G_{\mu\nu}=0 \textrm{ otherwise} ).
\end{align}
The solutions for the equations of motion should satisfy the null energy condition and allow the metric to be asymptotically hyperscaling violating.
They are given by
\begin{align}
\begin{split}
	F_{rt}
    	&= \rho_F\, e^{-\lambda_F \phi} r^{-\alpha(d+2)-z-d+1} \frac{r^{2\alpha}}{r^2 f(r)}
    	 	r^{2\alpha + 2z} f(r)\\
    	&= \rho_F\, e^{-\lambda_F \phi} r^{\alpha(2-d) + z - d - 1},
\end{split}
	\\
\begin{split}
	G_{rt}
    	&= \rho_G\, e^{-\lambda_G \phi} r^{\alpha(2-d) + z - d - 1},
\end{split}
	\\
	e^\phi &= e^{\phi_0} r^{\sqrt{2d(\alpha + 1)(\alpha + z - 1)}},
\end{align}
together with
\begin{align}
    f(r) = 1 - r^{-d\alpha - z - d}.
\end{align}
The coupling constants in the theory are not arbitrary but 
\begin{align}
	\gamma &= \frac{-2\alpha}{\sqrt{2d (\alpha + 1)(\alpha + z - 1)}},
	\\
    \lambda_F &= - \frac{2\alpha (d - 1) + 2d}{\sqrt{2d (\alpha + 1)(\alpha + z - 1)}},
    \\
    \lambda_G &= \sqrt{\frac{2(\alpha + z - 1)}{d(\alpha + 1)}},
    \\
    (\rho_F)^2 &= \frac{2V_0(z-1)}{d\alpha+d+z-1} e^{-\sqrt{ \frac{2d(\alpha+1)}{\alpha+z-1} }\phi_0},
\end{align}
and $\rho_G$ is undetermined.
It is interpreted that $F$ is related to an anisotropic scaling and $G$ gives the charge potential to the black brane.

We would like to promote this model to the Einstein-Scalar-$U(1)\times SU(2)$ theory\cite{Herzog:2009ci}.
Since $G$ gives the chemical potential, we leave $F$ as it is and change $G$ to the field strength $G^a$ of the SU(2) field $B^a$.
The parameters $\lambda_F$ and $\lambda_G$ won't change \cite{Park:2016wch}, so we relabel them as $\lambda_U$ and $\lambda_{YM}$, respectively.
The SU(2) field is defined by
\begin{align}
    G^a_{MN} = \partial_M B^a_N - \partial_N B^a_M - \epsilon^{abc} B^b_M B^c_N,
\end{align}
with the equations of motion given by
\begin{align}
    \mathcal{Y}^{aN}
    \equiv
    \frac{1}{\sqrt{-g}} \partial_M \left( \sqrt{-g} e^{\lambda_{YM} \phi} G^{aMN} \right)
    + e^{\lambda_{YM} \phi} \epsilon^{abc} G^{bMN} B_M^c = 0.
\end{align}
Note that $G^a_{MN} \rightarrow g_{YM}^{-1} G^a_{MN}$ as $B^a_M \rightarrow g_{YM}^{-1} B^a_M$.
We choose an ansatz
\begin{align}
    B^a \tau^a &= b(r) \tau^3 dt + w(r) \tau^1 dx_1.
\end{align}
Then, the components of $G^a_{MN}$ are given by
\begin{alignat}{4}
    G^3_{rt}
    &= \partial_r B_t^3 - \partial_t B_r^3 - \epsilon^{312} B_r^1 B_t^2 - \epsilon^{321} B_r^2 B_t^1
    &&= \partial_r b(r),
    \\
    G^1_{rx}
    &= \partial_r B_x^1 - \partial_x B_r^1 - \epsilon^{123} B_r^2 B_x^3 - \epsilon^{132} B_r^3 B_x^2
    &&= \partial_r w(r),
    \\
    G^2_{tx} &= -\epsilon^{213} B_t^1 B_x^3 - \epsilon^{231} B_t^3 B_x^1 &&= -b(r) w(r).
\end{alignat}
Let $u = r/r_0$, where $r_0 = 1$ is the bulk coordinate at which the horizon lies.
The equations of motion is explicitly written as
\begin{alignat}{2}
    \mathcal{Y}^{3t}
    &=0
    = u^{-\alpha(d + 2) - z - d + 1} \partial_u \left[ u^{d(\alpha + 1) + z - 1} \partial_u b(u) \right] - \frac{u^{-2(\alpha + 2)}}{1 - u^{-d\alpha - z - d}} b(u) w^2(u),
    \\
    \mathcal{Y}^{1x}
    &=0
    = u^{-\alpha(d + 2) - z - d + 1} \partial_u \left[
    u^{d\alpha + d + 3z - 3} \left( 1 - u^{-d\alpha - z - d} \right) \partial_u w(u) \right]
    + \frac{u^{-2\alpha - 4}}{1 - u^{-d\alpha - z - d}} b^2(u) w(u).
\end{alignat}
There is a special condition under which solutions for $b(u)$ and $w(u)$ can be obtained algebraically.
First, we restrict $d \alpha + d + z = 4$.
Then, the Null Energy condition yields the following inequalities:
\begin{align}
    (\alpha + 1)(\alpha + z - 1) &\ge 0\\
    \frac{3}{d}(\frac{3}{d} - 1) &\ge 0.
\end{align}
It follows that the only possible values $d$ can have are 1, 2, and 3.
Here, we consider the $AdS_5$ space, for which $d = 3$.
In this case, $\mathcal{Y}^{3t}$ and $\mathcal{Y}^{1x}$ can be written as
\begin{alignat}{2}
    \mathcal{Y}^{3t}
    &=0
    = u^{-2(\alpha + 2)} \left[ u\, \partial_u \left( u^3 \partial_u b(u) \right) - \frac{b(u) w^2(u)}{1 - u^{-4}} \right],
    \\
    \mathcal{Y}^{1x}
    &=0
    = u^{-2(\alpha + 2)} \left\{ u\, \partial_u \left[
    u^{2z+1} (1 - u^{-4}) \partial_u w(u) \right]
    + \frac{b^2(u) w(u)}{1 - u^{-4}} \right\}.
\end{alignat}
If we set $2z + 1 = 3$, we reach the desired condition.
We compute the solutions for $b(u)$ and $w(u)$ analytically, with $z=1$, $\alpha=0$, $d=3$.

First we introduce a book-keeping parameter $\varepsilon$ to write $b(u)$ and $w(u)$ as
\begin{align}
    w(u) = \sum_{n=0} \varepsilon^{2n+1} w_{2n+1}(u)
    \qquad \textrm{and} \qquad
    b(u) = \sum_{n=0} \varepsilon^{2n} b_{2n}(u).
\end{align}
The parameter $\varepsilon$ is going to be considered to be small, and the corrections will be obtained order by order.
We start from $b_0(u)$ and $w_1(u)$, which are given by
\begin{align}
    b_0(u) &= \mu \left( 1 - u^{-2} \right),\\
    w_1(u) &= \frac{u^2}{(1+u^2)^2},
\end{align}
where $\mu$ is the chemical potential.
The chemical potential $\mu$ is defined by
\begin{align}
    \mu = \lim_{u \rightarrow \infty} b(u) - \lim_{u \rightarrow 1} b(u).
\end{align}
Now we put $w(u)=\varepsilon w_1(u)$ into $\mathcal{Y}^{3t}$ to obtain $b_2(u)$.
It is given by
\begin{align}
    b_2(u) = \frac{1 + u^2(2 - 18c_1) - 6c_1 - 18u^4c_1 - 6u^6c_1}{12u^2(1+u^2)^3} + c_2,
\end{align}
where $c_1$ and $c_2$ are constants of integration.
These constants are going to be determined by requiring of $b(u)$ and $w(u)$ to vanish at horizon ($u = 1$).
Before that, we may write $c_1$ and $c_2$ in terms of $\delta \mu_2$, the second-order correction of the chemical potential $\mu$.
By the definition of $\mu$, it should be corrected in order of $\varepsilon$ as $b(u)$ is corrected.
We have
\begin{align}
    \delta\mu_2
    &= \lim_{u \rightarrow \infty} b_2(u) - \lim_{u \rightarrow 1} b_2(u) = \frac{1 - 16c_1}{32},\\
    c_1 &= \frac{1 + 32\delta\mu_1}{16}.
\end{align}
The vanishing condition of $b_2(u)$ at horizon gives
\begin{align}
    \lim_{u \rightarrow 1} b_2(u) &= c_2 + \frac{3 - 3(1 + 32\delta\mu_1)}{96} = 0,\\
    c_2 &= \delta\mu_1.
\end{align}
Hence $b_2(u)$ is written as
\begin{align}
    b_2(u) = \delta \mu_1 + \frac{5 + 7u^2 - 9u^4 - 3u^6 -96(1+u^2)^3 \delta\mu_1}{96u^2(1+u^2)^3}.
\end{align}
Now, the equation for $w_3(u)$ is written as
\begin{align}
    0 &=
    u\ \partial_u \left[ u^3(1-u^{-4}) \varepsilon^3 \partial_u w_3(u) \right]
    + \frac{(b_0(u) + \varepsilon^2 b_2(u))^2(\varepsilon w_1(u) + \varepsilon^3 w_3(u))}{1-u^{-4}}
\end{align}
The solution is given by
\begin{align}
   w_3(u)
   = \frac{1}{40320 \left(u^2+1\right)^5}
   &\left[ 2 u^2 \left( 20160 c_1 \left(u^2+1\right)^3-9 u^2 \left( 41u^2+91 \right)-331 \right) \right.
   \\
   & +3 u^2 \left(u^2+1\right)^3 \left\{ (-53760c_2+6720 \delta\mu_1 -71) \log\left( 1-u^2 \right) \right.
   \\
   &\left. +16 (3360c_2-13) \log (u)+35 (5-192 \delta\mu_1 ) \log
   \left(u^2+1\right) \right\}
   \\
   &\left. -20160 c_2 \left(u^4+1\right)
   \left(u^2+1\right)^3+78 \right].
\end{align}
The constant $c_1$ is the coefficient of $u^2/(u^2+1)^2$.
Because it is the zero mode, we may fix $c_1 = 0$.
For $c_2$, we check the regularity of the solution at horizon.
There are logarithmic terms which diverge at $u = 1$.
We may eliminate such terms by choosing
\begin{align}
    c_2 = \frac{\delta \mu_1 }{8}-\frac{71}{53760}
\end{align}
Lastly, the condition $w_3(u) = 0$ as $u\rightarrow\infty$ results in
\begin{align}
    \delta\mu_1 = \frac{71}{6720}.
\end{align}
Finally we have
\begin{align}
\begin{split}
    w_3(u) = -\frac{1}{40320 \left(u^2+1\right)^5}
    &\left[738 u^6+1638 u^4+662 u^2+624 \left(u^2+1\right)^3 u^2 \log
   (u)-312 u^2 \log \left(u^2+1\right) \right. \\
   &\left. -312 u^8 \log\left(u^2+1\right)-936 u^6 \log \left(u^2+1\right)-936 u^4 \log
   \left(u^2+1\right)-78
    \right]
\end{split}
\end{align}

\section{Sturm-Liouville problem}
Let us explain this Sturm-Liouville problem in more detail. Once one plugs the perturbative expansion of the solutions $w(u)$ and $b(u)$ given in \eqref{perturbative_wb}, the zeroth order equation in $\varepsilon$ is given by
\begin{align}
    \mathcal{Y}^{3t} = 0 = u^{-2(\alpha + 2)} [u \partial_u (u^3 \partial_u b(u))]
\end{align}
Its solution is
\begin{align}
    b(u) = \sqrt{3} \mu (1 - u^{-2})
\end{align}
where $\mu$ is dimensionless chemical potential (more precisely $\sqrt{3} \mu$ is chemical potential but our terminology is that $\mu$ is chemical potetial). Now we turn to 1st order equation in $\varepsilon$ is from $\mathcal{Y}^{1x}$ equation.
\begin{align}
    \mathcal{Y}^{1x} = 0 = u^{-2(\alpha + 2)} \left\{ u \partial_u [u^3 (1 - u^{-4}) \partial_u w_1(u)] + \frac{3 \mu^2 (1 - u^{-2})^2 w_1(u)}{1 - u^{-4}} \right\}
\end{align}
This is Sturm-Liouville problem. Since the solution $w_1$ exists only when the value of chemical potential $\mu$ is chosen appropriately.\\
The answer is in precedent researches[]. This solution is obtained in the background spacetime of asymptotically AdS$_5$. The gravity model for this is Einstein-SU(2) Yang-Mills theory. We note that this solution can also be obtained when we take $D = d + 2 = 5$, $\alpha = 0$, $z = 1$, where our scalar field $\phi$, and U(1) gauge fields become trivial i.e. $\phi = \phi_0$, $A = 0$.\\
The authors in these papers that the solution of $w_1(u)$ exists when $\sqrt{3} \mu = 4$, and the form of the solution of $w_1(u)$ is given by
\begin{align}
    w_1(u) = \frac{u^2}{(1 + u^2)^2}
\end{align}
The analytic solutions for subleading corrections as $b_2$ is given as below.
\begin{align}
    b_2(u) = \frac{71}{6720} \left( 1 - \frac{1}{u^2} \right) + \frac{5 + 7 u^2 - 9 u^4 - 3 u^6}{96 u^2 (1 + u^2)^3}
\end{align}
We stress that analytic solutions also exists in 3- and 4-dimensional bulk gravity theories, where the asymptotic spacetime becomes hyperscaling violation geometry. The analytic solutions can be obtained by imposing several conditions.


\begin{thebibliography}{9}

\bibitem{Aharony:1999ti}
O.~Aharony, S.~S.~Gubser, J.~M.~Maldacena, H.~Ooguri and Y.~Oz,
Phys. Rept. \textbf{323}, 183-386 (2000)
doi:10.1016/S0370-1573(99)00083-6
[arXiv:hep-th/9905111 [hep-th]].

\bibitem{Maldacena:1997re}
J.~M.~Maldacena,
Adv. Theor. Math. Phys. \textbf{2}, 231-252 (1998)
doi:10.1023/A:1026654312961
[arXiv:hep-th/9711200 [hep-th]].

\bibitem{Oh:2011wpl}
J.~H.~Oh,``Gauge-gravity Duality and its Applications to Cosmology and Fluid Dynamics''.

\bibitem{Kovtun:2004de}
P.~Kovtun, D.~T.~Son and A.~O.~Starinets,
Phys. Rev. Lett. \textbf{94}, 111601 (2005)
doi:10.1103/PhysRevLett.94.111601
[arXiv:hep-th/0405231 [hep-th]].

\bibitem{Bhattacharyya:2007vjd}
S.~Bhattacharyya, V.~E.~Hubeny, S.~Minwalla and M.~Rangamani,
JHEP \textbf{02}, 045 (2008)
doi:10.1088/1126-6708/2008/02/045
[arXiv:0712.2456 [hep-th]].

\bibitem{Bhattacharyya:2008ji}
S.~Bhattacharyya, R.~Loganayagam, S.~Minwalla, S.~Nampuri, S.~P.~Trivedi and S.~R.~Wadia,
JHEP \textbf{02}, 018 (2009)
doi:10.1088/1126-6708/2009/02/018
[arXiv:0806.0006 [hep-th]].

\bibitem{Benincasa:2006fu}
P.~Benincasa, A.~Buchel and R.~Naryshkin,
Phys. Lett. B \textbf{645}, 309-313 (2007)
doi:10.1016/j.physletb.2006.12.030
[arXiv:hep-th/0610145 [hep-th]].

\bibitem{Iqbal:2008by}
N.~Iqbal and H.~Liu,
Phys. Rev. D \textbf{79}, 025023 (2009)
doi:10.1103/PhysRevD.79.025023
[arXiv:0809.3808 [hep-th]].

\bibitem{Hartnoll:2009sz}
S.~A.~Hartnoll,
Class. Quant. Grav. \textbf{26}, 224002 (2009)
doi:10.1088/0264-9381/26/22/224002
[arXiv:0903.3246 [hep-th]].

\bibitem{Herzog:2009xv}
C.~P.~Herzog,
J. Phys. A \textbf{42}, 343001 (2009)
doi:10.1088/1751-8113/42/34/343001
[arXiv:0904.1975 [hep-th]].

\bibitem{Horowitz:2010gk}
G.~T.~Horowitz,
Lect. Notes Phys. \textbf{828}, 313-347 (2011)
doi:10.1007/978-3-642-04864-7\_10
[arXiv:1002.1722 [hep-th]].



\bibitem{Alishahiha:2012qu}
M.~Alishahiha, E.~O Colgain and H.~Yavartanoo,
JHEP \textbf{11}, 137 (2012)
doi:10.1007/JHEP11(2012)137
[arXiv:1209.3946 [hep-th]].


\bibitem{Gubser:2008px}
S.~S.~Gubser,
Phys. Rev. D \textbf{78}, 065034 (2008)
doi:10.1103/PhysRevD.78.065034
[arXiv:0801.2977 [hep-th]].

\bibitem{Gubser:2008zu}
S.~S.~Gubser,
Phys. Rev. Lett. \textbf{101}, 191601 (2008)
doi:10.1103/PhysRevLett.101.191601
[arXiv:0803.3483 [hep-th]].

\bibitem{Hartnoll:2008kx}
S.~A.~Hartnoll, C.~P.~Herzog and G.~T.~Horowitz,
JHEP \textbf{12}, 015 (2008)
doi:10.1088/1126-6708/2008/12/015
[arXiv:0810.1563 [hep-th]].

\bibitem{Herzog:2009ci}
C.~P.~Herzog and S.~S.~Pufu,
JHEP \textbf{04}, 126 (2009)
doi:10.1088/1126-6708/2009/04/126
[arXiv:0902.0409 [hep-th]].

\bibitem{Basu:2008bh}
P.~Basu, J.~He, A.~Mukherjee and H.~H.~Shieh,
JHEP \textbf{11}, 070 (2009)
doi:10.1088/1126-6708/2009/11/070
[arXiv:0810.3970 [hep-th]].

\bibitem{Ryu:2006bv}
S.~Ryu and T.~Takayanagi,
Phys. Rev. Lett. \textbf{96}, 181602 (2006)
doi:10.1103/PhysRevLett.96.181602
[arXiv:hep-th/0603001 [hep-th]].

\bibitem{Ryu:2006ef}
S.~Ryu and T.~Takayanagi,
JHEP \textbf{08}, 045 (2006)
doi:10.1088/1126-6708/2006/08/045
[arXiv:hep-th/0605073 [hep-th]].

\bibitem{Erdmenger:2010xm}
J.~Erdmenger, P.~Kerner and H.~Zeller,
Phys. Lett. B \textbf{699}, 301-304 (2011)
doi:10.1016/j.physletb.2011.04.009
[arXiv:1011.5912 [hep-th]].

\bibitem{Basu:2011tt}
P.~Basu and J.~H.~Oh,
JHEP \textbf{07}, 106 (2012)
doi:10.1007/JHEP07(2012)106
[arXiv:1109.4592 [hep-th]].

\bibitem{Oh:2012zu}
J.~H.~Oh,
JHEP \textbf{06}, 103 (2012)
doi:10.1007/JHEP06(2012)103
[arXiv:1201.5605 [hep-th]].

\bibitem{Park:2016wch}
M.~Park, J.~Park and J.~H.~Oh,
Eur. Phys. J. C \textbf{77}, no.11, 810 (2017)
doi:10.1140/epjc/s10052-017-5382-8
[arXiv:1609.08241 [hep-th]].

\bibitem{Park:2022oek}
C.~Park, G.~Kim, J.~s.~Chae and J.~H.~Oh,
JHEP \textbf{02}, 182 (2023)
doi:10.1007/JHEP02(2023)182
[arXiv:2210.08919 [hep-th]].

\bibitem{Solodukhin:2008dh}
S.~N.~Solodukhin,
Phys. Lett. B \textbf{665}, 305-309 (2008)
doi:10.1016/j.physletb.2008.05.071
[arXiv:0802.3117 [hep-th]].

\bibitem{Hung:2011xb}
L.~Y.~Hung, R.~C.~Myers and M.~Smolkin,
JHEP \textbf{04}, 025 (2011)
doi:10.1007/JHEP04(2011)025
[arXiv:1101.5813 [hep-th]].

\bibitem{Casini:2011kv}
H.~Casini, M.~Huerta and R.~C.~Myers,
JHEP \textbf{05}, 036 (2011)
doi:10.1007/JHEP05(2011)036
[arXiv:1102.0440 [hep-th]].

\bibitem{Bhattacharya:2012mi}
J.~Bhattacharya, M.~Nozaki, T.~Takayanagi and T.~Ugajin,
Phys. Rev. Lett. \textbf{110}, no.9, 091602 (2013)
doi:10.1103/PhysRevLett.110.091602
[arXiv:1212.1164 [hep-th]].

\bibitem{Bianchi:2012ev}
E.~Bianchi and R.~C.~Myers,
Class. Quant. Grav. \textbf{31}, 214002 (2014)
doi:10.1088/0264-9381/31/21/214002
[arXiv:1212.5183 [hep-th]].

\bibitem{Nozaki:2013vta}
M.~Nozaki, T.~Numasawa, A.~Prudenziati and T.~Takayanagi,
Phys. Rev. D \textbf{88}, no.2, 026012 (2013)
doi:10.1103/PhysRevD.88.026012
[arXiv:1304.7100 [hep-th]].

\bibitem{Allahbakhshi:2013rda}
D.~Allahbakhshi, M.~Alishahiha and A.~Naseh,
JHEP \textbf{08}, 102 (2013)
doi:10.1007/JHEP08(2013)102
[arXiv:1305.2728 [hep-th]].

\bibitem{Wong:2013gua}
G.~Wong, I.~Klich, L.~A.~Pando Zayas and D.~Vaman,
JHEP \textbf{12}, 020 (2013)
doi:10.1007/JHEP12(2013)020
[arXiv:1305.3291 [hep-th]].

\bibitem{Momeni:2015vka}
D.~Momeni, M.~Raza, H.~Gholizade and R.~Myrzakulov,
Int. J. Theor. Phys. \textbf{55}, no.11, 4751-4758 (2016)
doi:10.1007/s10773-016-3098-4
[arXiv:1505.00215 [hep-th]].

\bibitem{Park:2012lzs}
C.~Park,
Adv. High Energy Phys. \textbf{2013}, 389541 (2013)
doi:10.1155/2013/389541
[arXiv:1209.0842 [hep-th]].

\bibitem{Park:2015hcz}
C.~Park,
Phys. Rev. D \textbf{93}, no.8, 086003 (2016)
doi:10.1103/PhysRevD.93.086003
[arXiv:1511.02288 [hep-th]].

\bibitem{Kim:2016jwu}
K.~S.~Kim and C.~Park,
Phys. Rev. D \textbf{95}, no.10, 106007 (2017)
doi:10.1103/PhysRevD.95.106007
[arXiv:1610.07266 [hep-th]].

\bibitem{Jeong:2022zea}
H.~S.~Jeong, K.~Y.~Kim and Y.~W.~Sun,
JHEP \textbf{06}, 078 (2022)
doi:10.1007/JHEP06(2022)078
[arXiv:2203.07612 [hep-th]].

\bibitem{Hung:2011nu}
L.~Y.~Hung, R.~C.~Myers, M.~Smolkin and A.~Yale,
JHEP \textbf{12}, 047 (2011)
doi:10.1007/JHEP12(2011)047
[arXiv:1110.1084 [hep-th]].

\bibitem{Huerta:2011qi}
M.~Huerta,
Phys. Lett. B \textbf{710}, 691-696 (2012)
doi:10.1016/j.physletb.2012.03.044
[arXiv:1112.1277 [hep-th]].

\bibitem{Banerjee:2015tia}
S.~Banerjee, Y.~Nakaguchi and T.~Nishioka,
JHEP \textbf{03}, 048 (2016)
doi:10.1007/JHEP03(2016)048
[arXiv:1508.00979 [hep-th]].

\bibitem{Hertzberg:2010uv}
M.~P.~Hertzberg and F.~Wilczek,
Phys. Rev. Lett. \textbf{106}, 050404 (2011)
doi:10.1103/PhysRevLett.106.050404
[arXiv:1007.0993 [hep-th]].


\bibitem{Lewkowycz:2012qr}
A.~Lewkowycz, R.~C.~Myers and M.~Smolkin,
JHEP \textbf{04}, 017 (2013)
doi:10.1007/JHEP04(2013)017
[arXiv:1210.6858 [hep-th]].

\bibitem{Rosenhaus:2014woa}
V.~Rosenhaus and M.~Smolkin,
JHEP \textbf{12}, 179 (2014)
doi:10.1007/JHEP12(2014)179
[arXiv:1403.3733 [hep-th]].

\bibitem{Rosenhaus:2014zza}
V.~Rosenhaus and M.~Smolkin,
JHEP \textbf{02}, 015 (2015)
doi:10.1007/JHEP02(2015)015
[arXiv:1410.6530 [hep-th]].

\bibitem{Park:2015dia}
C.~Park,
Phys. Rev. D \textbf{92}, no.12, 126013 (2015)
doi:10.1103/PhysRevD.92.126013
[arXiv:1505.03951 [hep-th]].

\bibitem{Arias:2012py}
R.~E.~Arias and I.~S.~Landea,
JHEP \textbf{01}, 157 (2013)
doi:10.1007/JHEP01(2013)157
[arXiv:1210.6823 [hep-th]].

\bibitem{Cai:2012nm}
R.~G.~Cai, S.~He, L.~Li and Y.~L.~Zhang,
JHEP \textbf{07}, 027 (2012)
doi:10.1007/JHEP07(2012)027
[arXiv:1204.5962 [hep-th]].

\bibitem{Srivastav:2019tkr}
A.~Srivastav, D.~Ghorai and S.~Gangopadhyay,
Eur. Phys. J. C \textbf{80}, no.3, 219 (2020)
doi:10.1140/epjc/s10052-020-7769-1
[arXiv:1911.08786 [hep-th]].

\bibitem{Kanitscheider:2008kd}
I.~Kanitscheider, K.~Skenderis and M.~Taylor,
JHEP \textbf{09}, 094 (2008)
doi:10.1088/1126-6708/2008/09/094
[arXiv:0807.3324 [hep-th]].

\bibitem{Kanitscheider:2009as}
I.~Kanitscheider and K.~Skenderis,
JHEP \textbf{04}, 062 (2009)
doi:10.1088/1126-6708/2009/04/062
[arXiv:0901.1487 [hep-th]].


\bibitem{Balasubramanian:1999re}
V.~Balasubramanian and P.~Kraus,
Commun. Math. Phys. \textbf{208}, 413-428 (1999)
doi:10.1007/s002200050764
[arXiv:hep-th/9902121 [hep-th]].
\end{thebibliography}
\end{document}